\documentclass[a4paper]{ar-1col}
\usepackage[numbers]{natbib}

\jname{Ann.~Rev.~Nuc.~Part.~Sci.}
\jvol{66}
\jyear{2016}
\doi{10.1146/(DOI)}

\def \lesssim {\mathrel{\vcenter
     {\offinterlineskip \hbox{$<$}\hbox{$\sim$}}}}
\def \gtrsim {\mathrel{\vcenter
     {\offinterlineskip \hbox{$>$}\hbox{$\sim$}}}}

\begin{document}

\markboth{Janka, Melson, \& Summa}{Physics of Core-Collapse Supernovae
in Three Dimensions}

\title{Physics of Core-Collapse Supernovae in Three Dimensions:\\ 
a Sneak Preview}

\author{Hans-Thomas Janka,$^1$ Tobias Melson,$^{1,2}$ and 
Alexander Summa$^1$
\affil{$^1$Max Planck Institute for Astrophysics, Karl-Schwarzschild-Str.~1,
85748 Garching, Germany; email: thj@mpa-garching.mpg,de}
\affil{$^2$Physik Department, Technische Universit\"at M\"unchen, 
James-Franck-Str.~1, 85748 Garching, Germany}
}

\firstpagenote{First page note to print below DOI/copyright line.}

\begin{abstract}
Nonspherical mass motions are a
generic feature of core-collapse supernovae, and 
hydrodynamic instabilities play a crucial role for 
the explosion mechanism. First successful neutrino-driven 
explosions could be obtained with self-consistent, first-principle
simulations in three spatial dimensions (3D). But 3D
models tend to be less prone to explosion than corresponding
axisymmetric (2D) ones. This has been
explained by 3D turbulence leading to energy cascading from large
to small spatial scales, inversely to the 2D case, thus disfavoring
the growth of buoyant plumes on the largest scales. Unless the
inertia to explode
simply reflects a lack of sufficient resolution in relevant regions,
it suggests that some important aspect may still be missing
for robust and sufficiently energetic neutrino-powered explosions.
Such deficits could be
associated with progenitor properties like rotation,
magnetic fields or pre-collapse perturbations, or with microphysics
that could lead to an enhancement of neutrino heating behind the
shock. 3D simulations have also revealed new phenomena that are not
present in 2D, for example spiral modes of the standing accretion
shock instability (SASI) and a stunning dipolar lepton-emission 
self-sustained asymmetry (LESA). Both impose time- and
direction-dependent variations on the detectable neutrino signal.
The understanding of these effects and of their consequences is still
in its infancy.
\end{abstract}

\maketitle

\tableofcontents

\section{INTRODUCTION:\\ NEUTRINO-DRIVEN EXPLOSIONS FROM A 
BROADER PERSPECTIVE}

A large variety of observational aspects indicates the 
importance of multidimensional effects in core-collapse 
supernovae. Width and smoothness of the light-curve maximum,
Doppler shifting and broadening as well as sub-structures of 
spectral lines, early emergence of X-ray and 
gamma-ray emission, and significant levels of polarization
in the case of Supernova~1987A and other well observed
supernovae testify the presence of large-scale asphericity
and radial mixing processes
\cite[for reviews, see][]{Arnett1989,Hillebrandt1989,Maeda2008}.
Also the gaseous remnants of supernovae exhibit global
deformation and inhomogeneities as morphological
fingerprints of explosion asymmetries. Clumpiness, 
filaments, and directional variations of the elemental 
composition are interpreted as consequences and relics
of explosion asymmetries whose origin could be linked to
the earliest moments of the explosion
\cite[e.g.,][]{Aschenbach1995,DeLaney2010,Milisavljevic2013,Milisavljevic2015,Grefenstette2014,Boggs2015}.
Measured natal kick velocities of young neutron stars,
which cannot be explained by the orbital velocities of
disrupted binary systems, provide another empirical hint
to considerable asymmetries that are present already in
the earliest phase of the 
explosion~\cite[e.g.,][]{Cordes1998,Laietal2001,Arzoumanian2002,Hobbsetal2005}.

Numerical simulations also demonstrate that multidimensionality
plays a crucial role to explain
why massive stars blow up
\cite[for recent reviews on this subject, see][]{Janka2012,Janka2012b,Kotake2012,Kotake2012a,Burrows2013,Foglizzo2015}.
Hydrodynamical instabilities in collapsing stellar cores
like convection and the standing accretion shock instability 
\cite[SASI;][]{Foglizzo2001,Foglizzo2002,Blondinetal2003} have
been shown to grow even from small initial perturbations on 
time scales relevant for the supernova mechanism. They
trigger asymmetries and nonradial mass motions on large
scales and create turbulent flows on small scales. Also 
rotation and magnetic fields can have a decisive influence
on the development of the explosion, because the initial
fields can be strongly amplified in the collapsing stellar
core by compression, winding in shear layers, and in particular
by the magnetorotational instability in differentially rotating
regions \cite[e.g.,][]{Akiyamaetal2003}.

Multidimensional effects are not just a by-product or
side-effect of the explosion, they are essential for the 
success of the mechanism. The presently known and accepted
``standard'' physics ---i.e., pre-collapse
conditions from stellar evolution calculations,
transport opacities for active neutrinos, nuclear reaction
rates and the neutron-star equation of state,
relativistic plasma dynamics and general relativistic gravity---
do not facilitate successful explosions in spherically 
symmetric (``one-dimensional'', 1D) simulations except for
stars close to the low-mass end of supernova progenitors
\cite[for details, see][]{Janka2012}.
This conclusion was unavoidable after elaborate, energy-
and velocity-dependent three-flavor neutrino transport
based on the direct solution of the Boltzmann
equation \cite{Liebendoerfer2001,Liebendoerfer2004} as well as
iterative solvers of the two-moment equations (i.e.\ neutrino
energy and momentum equations) coupled to a Boltzmann closure
\cite{Burrows2000,Rampp2002}
had become available and did not bring success to the models.

\begin{textbox}[b]
\section{FLAVOR OSCILLATIONS OF ACTIVE NEUTRINOS}
Flavor oscillations of the three active
neutrino flavors are not self-consistently included
in present supernova models. 
For solar and atmospheric mixing parameters,
flavor changes induced by the matter background according to
the Mikheyev-Smirnov-Wolfenstein \cite{Mikheyev1985,Wolfenstein1978}
effect take place in the stellar mantle and envelope
(at densities $\lesssim$10$^4$\,g\,cm$^{-3}$)
and are suppressed in the dense medium of the
supernova core \cite{Wolfenstein1979,Hannestad2000}. The
situation is less clear for collective flavor transformations
caused by neutrino-neutrino interactions. In view of the still
incomplete understanding of this extremely complex,
highly nonlinear phenomenon, it
cannot be excluded that neutrinos propagating
through the dominant neutrino background outside of the
neutrinosphere might change their flavor identity
\cite[for a status report, see][]{Mirizzi2016}.
Oscillations of active neutrinos outside of the neutrinosphere
are important for the detectable neutrino signal and may have
an impact on supernova nucleosynthesis, but it is unlikely
that they have a strong influence on the explosion.
Since in present supernova models the individual luminosities 
of muon and tau neutrinos and antineutrinos ($\nu_\mu$, $\nu_\tau$,
$\bar\nu_\mu$, $\bar\nu_\tau$, collectively denoted as $\nu_x$)
are roughly half of those of
electron neutrinos ($\nu_e$) and antineutrinos ($\bar\nu_e$) 
during the post-bounce accretion phase, and since the
heavy-lepton neutrino spectra
are only moderately harder, even a complete swap of 
$\nu_{\mu,\tau}$ to $\nu_e$ (or of the antineutrinos) just
outside of the neutrinosphere would not enhance the 
neutrino-energy deposition behind the stalled shock front.
\end{textbox}

Solving the riddle how supernovae explode therefore requires
a better understanding of multidimensional phenomena in the
collapsing stellar core. Self-consistent
multidimensional numerical simulations are indispensable
for that and have enabled considerable
progress over the past decade. Supplemented by semi-analytic 
analysis, toy models, parametric studies, and in the case of
the SASI even by laboratory
experiments~\cite{Foglizzo2012,Foglizzo2015}, such simulations
have advanced the field, providing us deeper insights 
into the nonlinear processes that could aid the revival of
the stalled bounce shock and shape the observable asymmetries
of supernova explosions.

Of course, astrophysical phenomena in nature involve three 
spatial dimensions and the ultimate aim of first-principle
modeling must therefore be 3D simulations. The increasing 
power of modern supercomputers and the development of 
highly parallelized numerical codes have made it possible
only very recently to perform such 3D simulations with grid-based
schemes including energy-dependent, three-flavor neutrino transport.
This progress could thus be achieved more than a decade after
the seminal work by C.~Fryer and collaborators, who obtained
neutrino-driven explosions in full-scale 3D stellar core-collapse
models with smoothed-particle
hydrodynamics (SPH) including gray, flux-limited neutrino diffusion
\cite{FryerWarren2002,FryerWarren2004,FryerYoung2007}. Their results
basically confirmed similar, earlier simulations in two
dimensions~\cite{Herantetal1994,Fryer1999} and represent
a modern manifestation of the original concepts developed by
Colgate \& White~\cite{Colgate1966}, Arnett~\cite{Arnett1966}, 
and Bethe \& Wilson~\cite{BetheWilson1985}. 
While these spearheading computations of the first 3D explosions
must unquestionably be considered as a breakthrough in 
supernova modeling, they still granted only a preliminary, diffuse
glimpse into the 3D world. The radical simplifications of the
neutrino transport and the severe deficiencies of resolution and
accuracy in the hydrodynamics treatment did not allow for strong
and reliable conclusions on the viability of the neutrino-driven
mechanism.

In this review we report the advances of 3D supernova modeling 
since these early steps and reflect our growing
understanding of the physics that determines the explosion
mechanism. Because the prompt bounce shock
fails to trigger explosions, the energy deposition by the enormously
intense neutrino flux radiated from the newly formed neutron star
appears as the most plausible explanation for the revival of the
stalled supernova shock. Neutrino heating may provide the 
engine that powers the far majority of
``normal'' supernovae with energies from less than
$\sim$10$^{50}$\,erg to roughly $2.5\times 10^{51}$\,erg
\cite{Fryer1999,Fryer2001,Ugliano2012,Ertl2016,Sukhbold2016}.

\begin{textbox}[b]
\section{ALTERNATIVES TO NEUTRINO-DRIVEN EXPLOSIONS}
In supernovae with energies higher than 
$\sim$2.5$\times 10^{51}$\,erg up to the ``hypernova'' regime, where the
explosions can reach several $10^{52}$\,erg for progenitor stars
above $\sim$20\,$M_\odot$, rapid rotation and the
efficient amplification of magnetic fields are likely to play
a crucial role. These explosions are probably driven by 
magnetohydrodynamic effects around highly magnetized, 
fast-spinning neutron stars (possibly associated with the 
observed ``magnetars'') or around rapidly rotating black
holes that accrete infalling stellar gas with high rates
from a surrounding torus threaded by strong magnetic fields
(see Janka 2012 for a review). 

Though recent multidimensional models seem to strengthen 
the case, the paradigm of the neutrino-driven mechanism is
far from being convincingly established or even proven.
It should not remain unmentioned, that it is still
questioned by some because of the
remaining problems of self-consistent models to yield robust 
explosions and to explain the observed energies of typical
supernovae by neutrino-energy deposition.
While these problems may disappear
once 3D models become more mature, one should certainly remain
open-minded as long as there is a lack of solid empirical
evidence for the neutrino mechanism. However, the suggested 
alternatives, e.g.\ the ``jittering-jet
mechanism''~\cite{Papish2011,Papish2014} and ``collapse-induced 
thermonuclear explosions''~\cite{Kushnir2015}, are based on ad-hoc
assumptions in tension with the presently established understanding
of stellar evolution and supernova dynamics. The involvement
of speculative ingredients and the missing self-consistency
are neither satisfactory nor convincing and  
place such suggestions on a level of sophistication 
far below the current state of the neutrino-driven mechanism.
\end{textbox}

\section{ADVANCING FROM TWO TO THREE DIMENSIONS}

In this review we focus on the physics
of the neutrino-driven mechanism and provide an update
on the developments in 3D modeling after the report
given by Janka~\cite{Janka2012}. Before doing so, this section 
will briefly summarize what we have learned from 2D 
simulations and why further progress needs 3D models.

\subsection{Status of 2D Modeling}
\label{sec:2D}

Most of the full-scale modeling of supernovae has been 
performed, and much of it is still performed, with the 
assumption of rotational symmetry around a chosen axis,
i.e., in two spatial dimensions (2D). Such simulations
with state-of-the-art neutrino transport lend support to
the viability of the neutrino-driven mechanism, although
the present results of different groups reveal important and 
unsatisfactory quantitative and qualitative differences, and
no general consensus has been achieved yet. 

\subsubsection{Results from Different Groups}

2D simulations by the Oak Ridge group with ``ray-by-ray-plus''
(RbR+) flux-limited diffusion for 12, 15, 20, and 25\,$M_\odot$
progenitors develop neutrino-driven explosions nearly 
simultaneously~\cite{Bruenn2013,Bruenn2016}. The supernova 
shock expands essentially identically in all cases, and shock
revival (measured by the time when the total energy of some
mass in the 
gain layer becomes positive) occurs after only $\sim$200\,ms
of post-bounce accretion. Because of the early onset of the
explosions these models possess a big mass in the gain layer
and therefore explode fairly strongly with energies
between 0.34 and 0.88\,B (1\,B = 1\,bethe = $10^{51}$\,erg),
which are still increasing at the end of the simulations. 
However, these ``Series B'' models seem to be too optimistic
and were (modestly) affected by a numerical 
deficiency~\cite{Lentz2015}.

In contrast, for the same stars the Garching simulations,
employing RbR+ two-moment transport
with Boltzmann closure, yield later explosions. The onset
times differ between the progenitors because the blast wave
expands only when the mass-accretion rate has dropped 
sufficiently~\cite{Summa2016}. The explosions are
therefore likely to be less energetic, but the simulations
had to be stopped due to small time steps long before 
the energies could saturate.

O'Connor and Couch~\cite{OConnor2016}, 
using a two-moment (M1) scheme
with algebraic closure relation for 2D neutrino transport (but
with a subset of neutrino interactions and without energy-bin 
coupling) found a dynamical behavior of the four progenitors
qualitatively very similar to the results of the Garching
group. Quantitative differences in details can probably be
traced back to differences in the modeling ingredients, which
demand systematic tests.

The Princeton group, applying 2D flux-limited neutrino diffusion 
diffusion~\cite{Dolence2015} and M1 transport~\cite{Skinner2016},
again for supernova runs of the same progenitors, 
did not obtain any explosions. However, they used a Newtonian 
potential, whereas all other groups made use of an approximation
of relativistic gravity. The quantitative importance of
relativistic effects had already been concluded from 
1D and multidimensional models in a variety of works 
\cite[e.g.,][]{Lentz2012,Mueller2012gr,Kuroda2012}.
O'Connor and Couch~\cite{OConnor2016} explicitly demonstrated that
their models did not explode when Newtonian gravity was employed.

Explosions in 2D for larger sets of progenitors were also reported
by Japanese groups~\cite{Suwa2016,Nakamura2015} for
Newtonian simulations with RbR neutrino transport based on the 
isotropic diffusion-source approximation 
\cite[IDSA;][]{Liebendoerfer2009}, by the Basel
group~\cite{Pan2016}
for Newtonian models with multidimensional IDSA (and a 
high-density equation of state different from the ones
used in all other works), and by the Garching group for
relativistic simulations with RbR+ two-moment transport and
Boltzmann closure \cite{Janka2012b,Mueller2012gr,Mueller2012sasi,Mueller2013,Mueller2014}. 

\subsubsection{Assessment}

Although it is assuring that the models in the 
subset of successful cases
exhibit similarities concerning their gross features 
\cite[with the explosions of Refs.][defining outliers in the optimistic direction]{Bruenn2013,Bruenn2016},
much of the agreement could just be 
incidental. Suwa et al.~\cite{Suwa2016}, for example, obtain an
explosion for the 12\,$M_\odot$ progenitor but not for the 15 
and 20\,$M_\odot$ stars investigated by the other groups mentioned
above, and Nakamura et al.~\cite{Nakamura2015} report an explosion of
an 11.2\,$M_\odot$ model with considerably faster shock expansion
and higher energy than found by the Garching team
\cite{MarekJanka2009,Janka2012b,Mueller2012gr}.
A detailed comparison of the published
results, even for the same progenitor stars, is hampered by the
fact that the simulations were not only performed with different
transport solvers and approximations but also with different 
hydrodynamic schemes
and computational grids, implying different (numerical) 
perturbations that seed the growth of instabilities. Moreover,
different resolutions, different sets of neutrino reactions 
with different simplifications for the opacities, different
equation-of-state descriptions in particular also in the 
low-density regime, and 
different gravity treatments enhance the difficulties 
of direct comparisons. Dedicated and coordinated code
tests involving the competing groups are highly desirable
and currently in preparation.

\begin{figure}[!]
\includegraphics[width=16cm]{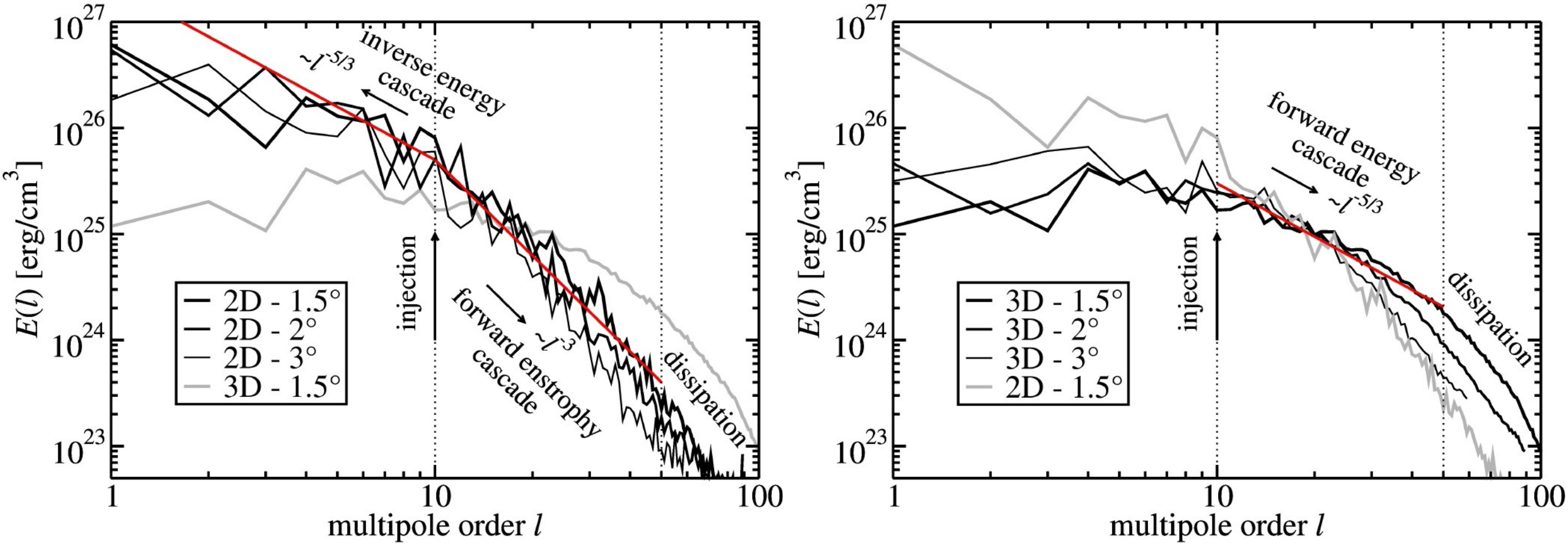}
\caption{
Turbulent energy spectra $E(l)$ as functions of the multipole order
$l$ for 2D and 3D simulations with different angular
resolution~\cite{Hanke2012}. The spectra are based on a decomposition
of the lateral velocity $v_\theta$ into spherical harmonics at a chosen
radius in the gain layer of a 15\,$M_\odot$ star during the post-bounce
accretion phase (400\,ms p.b.). The left panel shows 2D models
with different angular resolution (black, different thickness) and,
for comparison, the 3D model with the highest employed angular resolution 
(gray). The right panel displays 3D models with different angular
resolution (black lines, different thickness) 
and, for comparison, the 2D model with the highest employed
angular resolution (gray). The power-law dependences and the direction
of the energy and enstrophy (i.e., the squared vorticity of the velocity
field) cascades in the inertial range according to 
Kolmogorov turbulence theory~\cite{Kolmogorov1941,Landau1959} 
are indicated by red lines. 
The left vertical, dotted line (at $l\approx 10$) roughly marks the  
energy-injection scale, which corresponds to the typical scale of
growing convective plumes.
The right vertical, dotted line denotes the onset of dissipation by
numerical effects on small scales for the best displayed resolution.
At high resolution, the energy contained
in small-scale disturbances increases, as the dissipation range
moves to larger $l$.
In 3D (right panel), a power-law spectrum 
with $E(l)\propto l^{-5/3}$ develops in the inertial range at
intermediate multipole orders (or wavenumbers), whereas in 2D 
(left panel) the power-law dependence $E(l)\propto l^{-5/3}$
approximately holds for $l\lesssim 10$ in 2D as a result of
the reverse energy cascade~\cite{Kraichnan1967}. The energy
injected at $l\approx 10$ is therefore not transferred to the
dissipative range; only enstrophy is transported in a forward
cascade with a different power-law index ($E(l)\propto l^{-3}$).
The spectra of 2D simulations are strongly dominated by power in 
the lowest modes ($l \le 4$), associated with SASI activity and
large-scale buoyant plumes.}
\label{fig:turbulence}
\end{figure}

\subsection{Need of 3D Models}
\label{sec:need3D}

While the tension associated with discrepant results of 
different groups
is unsatisfactory, supernova theory and 2D simulations
have never been joined in a love
marriage but simply a partnership of convenience that was
enforced by the limitations of computing resources and the
constraints of numerical codes.
Two-dimensional modeling has played, and may 
continue to play, an important role for understanding basic
effects of nonradial flows and their feedback on the neutrino
emission and heating in self-consistent,
multidimensional supernova models.
The simplifying assumption of axisymmetry eases 
the modeling enormously by being computationally
much less demanding than 3D simulations. 
Currently, 3D calculations are at the extreme limit
of numerical feasibility because of their tremendous
requirements of computing resources, which strongly
increase with more elaborate neutrino treatment.
Therefore 2D calculations have so far been the only
way to perform resolution studies with full-scale supernova
models and to explore larger model sets for the dependence
of explosions on microphysics and progenitor properties.

However, important questions cannot be finally 
answered in two dimensions. How robust is the neutrino-driven 
mechanism? What are the uncertainties associated with
remaining approximations
and simplifications in the neutrino transport and microphysics? 
How important are amplitude and mode pattern of perturbations
in the pre-collapse core? When and how does rotation play a
role for the explosion, when do magnetic fields become 
dynamically relevant? Can neutrino-driven explosions explain
the observed asymmetries of supernovae and supernova remnants
as well as pulsar kicks and spins? 

Three-dimensional simulations are certainly
needed to connect models to real observations. Moreover,
they are indispensable to quantitatively assess the
dependence of the mechanism on hydrodynamic instabilities
for several reasons. First, in 3D
new phenomena and new modes of instability can come
into play that do not exist in 2D. For example, in addition
to axisymmetric ($m=0$) SASI sloshing motions in the 2D case,
3D simulations can develop ($m=1$) spiral SASI
modes, too \cite{BlondinMezzacappa2007,Iwakamietal2009,Fernandez2010,Hanke2013,Abdikamalov2015,Nakamura2014}.
Second, in agreement with theoretical
expectations~\cite{Kraichnan1967}, the energy spectrum of 
turbulent mass motions
was found to differ between 2D and 3D and the turbulent cascading 
to transport energy in opposite directions 
\cite[\textbf{Figure~\ref{fig:turbulence}};][]{Hanke2012,Couch2013,Dolence2013,Handy2014,Abdikamalov2015}.
Third, the imposed axisymmetry
constrains SASI-driven or convectively driven large-amplitude
shock oscillations to the axial direction, which was claimed
to lead to unphysical feedback effects on the neutrino emission,
in particular when the RbR approximation is 
applied~\cite{Skinner2016,Dolence2015}. Interestingly,
however, the differences in the dynamical evolution for models
with M1 and RbR transport reported by Skinner et al.~\cite{Skinner2016}
decreased when their simulations were performed with higher 
resolution. Moreover, these differences are in conflict with the
overall consistency between the models with M1 transport 
by O'Connor \& Couch~\cite{OConnor2016} and those by the Garching 
group~\cite{Summa2016}, which were computed with RbR+ transport.

\section{THE EXPLOSION MECHANISM IN THREE DIMENSIONS}

For the reasons discussed in Sect.~\ref{sec:need3D},
3D simulations are the long-desired, natural next step
in supernova modeling with grid-based hydrodynamics schemes.
First results have been published since 2010 and have already
led to interesting insights and even the discovery of new
phenomena.

\subsection{Parametric and Self-consistent 3D Modeling}

Similar to the spectrum of published 2D simulations 
reported in Sect.~\ref{sec:2D}, 3D calculations are
performed with a wide variety of hydrodynamics codes and
modeling strategies, using
different grids and resolutions, different gravity treatments,
different equations of state (ranging from simple ideal-gas
laws to state-of-the-art descriptions of the hadronic and 
leptonic plasma components in all relevant regimes of 
density, temperature, and composition), and different 
approximations for the neutrino treatment, neutrino 
opacities, and included neutrino reactions. All currently 
applied transport schemes involve approximations, because 
a rigorous, time-dependent solution of the Boltzmann
transport equation in six-dimensional phase space (with
three spatial coordinates and three momentum components)
is not feasible on the available supercomputers but will
require exaflop capability. 

The neutrino treatments in
time-dependent 3D models in the literature can be sorted
into three basic categories, differing in their degree
of sophistication, namely:
\begin{itemize}
\item 
No transport but only
schematic source terms for neutrino heating and/or cooling, 
sometimes coupled to a light-bulb assumption with a chosen,
fixed value of the neutrino luminosity 
\cite[e.g.,][]{Nordhausetal2010,Hanke2012,Dolence2013,Couch2013,Handy2014,Nakamura2014,Fernandez2015,Radice2016}.
\item
Crude approximations of neutrino transport like leakage
schemes with neutrino heating
\cite[e.g.,][]{Ott2013,Couch2013,Couch2014,Abdikamalov2015},
IDSA~\cite{Takiwaki2012,Takiwaki2014} or simple 
integrators of the gray \cite{Schecketal2006,Wongwathanaratetal2010,MuellerE2012,Wongwathanarat2013,Wongwathanarat2015}
or energy-dependent transport equations 
\cite{Mueller2015,Mueller20152d3d}.
\item
State-of-the-art transport based on 
energy-dependent solvers for flux-limited neutrino 
diffusion~\cite{Lentz2015}, for the two-moment equations
with Boltzmann closure 
\cite{Hanke2013,Tamborra2013,Tamborra2014a,Tamborra2014b,Melson2015a,Melson2015b},
or the M1 equations with analytic variable Eddington 
factor~\cite{Kuroda2012,Kuroda2015}.
\end{itemize}
All non-leakage transport treatments used in 
time-dependent 3D simulations so far have employed
the RbR(+) approximation for the directional variations
except the works by Kuroda et al.~\cite{Kuroda2012,Kuroda2015},
which, however, are severely 
limited in resolution and could cover only 
some 10 to 100\,ms after bounce.

Ignoring neutrinos completely or not describing 
neutrino transport implies a lack of self-consistency.
In particular the omission of feedback
effects of the hydrodynamics (e.g.\ of accretion or
rotation) on neutrino emission and heating can be problematic
and can seriously limit the conclusions that can be drawn.
In the following, such aspects of numerical modeling
will not be commented on unless they are relevant
for the results and associated discussion.

\subsection{The Importance of Nonradial Flows}
\label{sec:nonradflow}

The impact of multidimensional flows on the neutrino
mechanism has been investigated since the publications of
the first 2D models \cite{Herantetal1994,Burrowsetal1995,JankaMueller1996}.
Now with the focus on the recent 3D results, the discussion
of the question is 
still going on, which effects play a role and on which
level of importance.

\subsubsection{Growth of Convection and Buoyancy}

The negative entropy gradient created by neutrino heating 
can cause the onset of convective overturn in the postshock
layer, whose growth rate (inverse growth time scale)
is connected to the complex Brunt-V\"ais\"al\"a
frequency: 
$\omega_\mathrm{buoy} = \mathrm{Im}(\omega_\mathrm{BV}) > 0$
for convective instability. The inward motion of
the postshock accretion flow (with negative radial velocity
$v_r$), however, suppresses the linear growth of buoyant 
perturbations unless~\cite{Foglizzoetal2006}
\begin{equation}
\chi \equiv \int_{R_\mathrm{g}}^{R_\mathrm{s}}
\mathrm{d}r\,\frac{\omega_\mathrm{buoy}}{|v_r(r)|}
\sim \frac{\tau_\mathrm{adv}}{\tau_\mathrm{conv}}
\gtrsim 2...3 \,.
\label{eq:chiparameter}
\end{equation}
Here, all quantities ($\omega_\mathrm{buoy}$, $v_r$, the
gain radius $R_\mathrm{g}$ and shock radius $R_\mathrm{s}$)
have to be angle-averaged
as discussed, e.g., in Ref.~\cite{Fernandez2014},
and $\tau_\mathrm{adv}\sim
(R_\mathrm{s}-R_\mathrm{g})/\langle|v_r|\rangle_\mathrm{g}$
and $\tau_\mathrm{conv}\sim 
\langle\omega_\mathrm{buoy}^{-1}\rangle_\mathrm{g}$ 
are the advection time scale and convective growth time scale
in the gain layer, respectively (the angle brackets denote 
volume averages). The typical accretion velocity of the postshock
flow, $v_r$, scales with the infall velocity of the stellar
matter ahead of the shock, which is roughly given by the 
free-fall velocity at the shock: 
$v_r\sim\beta^{-1}v_\mathrm{ff}(R_\mathrm{s})
\approx -\beta^{-1}\sqrt{2GM_\mathrm{NS}/R_\mathrm{s}}$ with $\beta$
being the density jump in the shock and $M_\mathrm{NS}$ being
the mass providing the accelerating gravity field (approximated
by the neutron-star mass). The threshold
defined by $\chi \gtrsim 2...3$, however, can be circumvented
when the initial density perturbations (compared to the 
ambient density $\rho_0$) are in the nonlinear 
regime already:
\begin{equation}
\delta_\rho \equiv \frac{|\rho-\rho_0|}{\rho_0} \gtrsim
\frac{\langle|v_r|\rangle_\mathrm{g}}{\langle g\rangle_\mathrm{g}
\tau_\mathrm{adv}}\sim {\cal O}(1\%) \,,
\label{eq:nonlinear}
\end{equation}
where $\langle g\rangle_\mathrm{g}$ is the average value of 
the gravitational acceleration in the gain layer. In this
case a small-scale perturbation is able to
rise against the accretion flow. If the whole flow is perturbed, the
buoyant motions on small scales affect the situation globally and
can allow for the onset of convective overturn also on larger
scales~\cite{Schecketal2008}.

\begin{textbox}[t]
\section{STANDING ACCRETION SHOCK INSTABILITY (SASI)}
The SASI is a generic, nonradial instability of
stagnant accretion shocks that leads to large-scale
shock deformation with dipolar and quadrupolar ($l=1,\,2$)
spherical harmonics modes having the biggest growth rates.
In the nonlinear regime violent, large-amplitude
sloshing and spiral motions of the shock are a consequence
\cite[e.g.,][]{Blondinetal2003,BlondinMezzacappa2007,Iwakamietal2008,Iwakamietal2009,Schecketal2008}.
The oscillatory growth of the SASI from initial perturbations 
is caused by an advective-acoustic cycle in the cavity
between stalled shock and accreting proto-neutron star
\cite{Foglizzo2001,Foglizzo2002,Foglizzoetal2007,Schecketal2008,Guilet2012}.
The saturation
amplitude of the SASI is determined by dissipative flow 
effects like parasitic instabilities (e.g., 
secondary Kelvin-Helmholtz and Rayleigh-Taylor instability),
which extract energy from the large-scale coherent flow.
A laboratory experiment with a hydraulic jump in a shallow 
water flow is a physics analogue of the stalled accretion
shock that shares basic features with the SASI in supernova
cores 
\cite[apart from neutrino heating and associated buoyancy;][]{Foglizzo2012,Foglizzo2015}.
Symmetry breaking of $m = 1$ modes to spiral SASI modes requires
that the ratio of initial shock radius to neutron-star radius  
exceeds a critical value of about two~\cite{Kazeroni2016}. 
Angular momentum
separation and SASI induced explosion asymmetries can be
important to explain neutron-star 
kicks~\cite{Schecketal2004,Schecketal2006}
and spins~\cite{BlondinMezzacappa2007,Guilet2014}.
\end{textbox}

\subsubsection{Growth of the SASI}
\label{sec:sasigrowth}

The linear growth rate of SASI shock deformation modes
due to an amplifying advective-acoustic cycle in the
accretion flow between stalled accretion shock and 
neutron-star surface can be coined 
as~\cite{Foglizzoetal2007,Schecketal2008}:
\begin{equation}
\omega_\mathrm{SASI} = \frac{\ln|{\cal Q}|}{\tau_\mathrm{cyc}}\,.
\label{eq:omegasasi}
\end{equation}
Here, ${\cal Q}$ is the cycle efficiency and 
\begin{equation}
\tau_\mathrm{cyc} = \int_{R_0}^{R_\mathrm{s}} 
\frac{\mathrm{d}r}{|v_r(r)|} + 
\int_{R_0}^{R_\mathrm{s}} \frac{\mathrm{d}r}{c_\mathrm{s}(r)}
\label{eq:cycletime}
\end{equation}
the duration of the cycle as the sum of sound travel time (second,
sub-dominant term;
$c_\mathrm{s}(r)\gg |v_r(r)|$ is the local sound speed) and advection 
time between shock radius and a cycle-coupling
radius $R_0$, which is located in the flow-deceleration region 
between neutron-star radius ($R_\mathrm{NS}$) and gain radius.
According to Eq.~(\ref{eq:omegasasi}) the SASI growth rate scales 
inversely with the cycle period and SASI activity is expected to
be strongest during retraction phases of the accretion shock.

\subsubsection{Consequences for the Explosion}

Dimension is a key to the neutrino mechanism of core-collapse
supernova explosions~\cite{Nordhausetal2010}. So far,
however, it is not clear which key exactly fits into the keyhole.

\paragraph{Improvements Compared to 1D}
It is undisputed and supported by all modern simulations that
nonradial postshock flows enhance the neutrino-energy deposition
compared to the 1D case.
This can be concluded from higher heating rates ($\dot Q_\nu$) 
and higher heating efficiencies
[$\eta_\nu = \dot Q_\nu/(L_{\nu_e}+L_{\bar\nu_e})$;
$L_{\nu_i}$ is the luminosity of one neutrino species] 
in the gain layer. Convective buoyancy and SASI motions
push the shock to larger radii, thus increasing the mass
($M_\mathrm{g}$) in the gain layer, which can be
interpreted as a stretching of the effective advection time through
the gain layer, $\tau_\mathrm{adv}\approx M_\mathrm{g}/|\dot M|$
with $\dot M$ being the progenitor-specific mass-accretion
rate~\cite{Burasetal2006B,MarekJanka2009}, or as an increase
of the dwell time of matter in the gain 
layer~\cite{Thompsonetal2005,MurphyBurrows2008B}.

In a more detailed picture of the flows, accretion downdrafts
channel cool matter close to the gain radius, where efficient
neutrino-energy deposition takes place. Simultaneously, rising
high-entropy plumes carry the neutrino-heated gas outwards,
away from the gain radius. The corresponding expansion cooling
of this gas diminishes the energy loss by reemission of neutrinos.

\paragraph{3D versus 2D: No Consensus}
A highly controversial and still not completely settled
discussion was instigated by the question how nonradial flows
differ in 2D and 3D and what the corresponding consequences 
for the explosion mechanism could be.
In first, parametric (neutrino light bulb) simulations in 3D,
Nordhaus et al.~\cite{Nordhausetal2010} found considerably earlier 
explosions than in 2D, requiring $\sim$15--25\%
lower driving neutrino luminosity than in 2D. Also subsequent,
revised work by the Princeton 
group~\cite{Burrows2012,Dolence2013}
still showed (slightly) more favorable 
explosion conditions in 3D. These
results could not be reproduced by Hanke et al.~\cite{Hanke2012},
who saw little difference between 2D and 3D for their
standard resolution but easier explosions with higher
angular resolution in 2D and the opposite trend in 3D. 

The findings by Hanke et al.~\cite{Hanke2012} were confirmed by
self-consistent 3D simulations of the Garching group with 
high-fidelity neutrino transport~\cite{Hanke2013,Melson2015b}. 
They also received support by other groups: explosions in 3D occur
less readily than in 2D
\cite{Couch2013,Couch2014,Takiwaki2012,Takiwaki2014,Lentz2015}.
With better numerical resolution, Couch~\cite{Couch2013} and
Abdikamalov et al.~\cite{Abdikamalov2015} obtained a longer 
delay of the shock revival and higher values of the critical
neutrino luminosity for explosions in 3D, in
accordance with Hanke et al.~\cite{Hanke2012}.

Fernandez~\cite{Fernandez2015}, in support of 
Nakamura et al.~\cite{Nakamura2014} and
Iwakami et al.~\cite{Iwakami2014},  
observed that SASI-dominated explosions can
be obtained with up to $\sim$20\% lower driving luminosity in
3D than in 2D because of the ability of spiral modes to store
more nonradial kinetic energy than linear sloshing modes.
The magnitude of this difference, however, decreases with
increasing resolution. 

While all of these works were concerned with the question
which critical neutrino luminosity is needed to trigger
shock revival, Handy et al.\ (2014) approached
the problem from a different angle and fixed the final
explosion energy instead of the driving neutrino luminosity
in their parametric setups. They found that 3D models
require lower neutrino luminosities to produce equally
energetic explosions as 2D simulations and concluded
that the ``convective engine'' in their models is 4\% 
more efficient in 3D than in 2D. This result is not in
conflict with any of the other ones because the authors
explored a different question.

\paragraph{Controversial Resolution Effects in 2D}
In contrast to the Garching group 
\cite{Scheck2006,Hanke2012,Summa2016},
Couch~\cite{Couch2013} as well as Skinner et al.~\cite{Skinner2016}
observed a trend to later explosions also for 
better resolved 2D models. Couch \& Ott~\cite{Couch2015} speculated
that this discrepancy might be a consequence of massively
underresolved turbulence in the Garching models, while their
own simulations produce the correct behavior in the inertial
range of turbulence. However, it is equally well possible that
the disparate resolution dependence of the 2D simulations
is simply caused by differences of the numerical schemes.

The code used at Garching 
\cite[employing spherical polar coordinates or a
Yin-Yang grid;][]{WongwathanaratHammer2010}
retains perfect spherical symmetry for spherical initial
data. Therefore the growth of hydrodynamic instabilities
must be seeded by artificially imposed perturbations.
Adding numerical resolution in these simulations indeed
reduces the effects of numerical viscosity and thus can
be conducive for the development of hydrodynamic instabilities.
The trend to (slightly) earlier 2D explosions with higher
angular resolution persists all the way from 3$^\circ$ down
to 0.5$^\circ$ bin width~\cite{Hanke2012,Summa2016} 
with signs of convergence between
1$^\circ$ and 0.5$^\circ$~\cite{Scheck2006}. This finding
can well mean that the flow dynamics that supports the
explosion is not fully described by the concept of 
turbulence. An inversion of the trend was obtained for 
higher radial resolution by Hanke et al.~\cite{Hanke2012}, however,
not because of a better representation of turbulence as claimed
by Couch \& Ott~\cite{Couch2015}, but because of a resolution
sensitivity connected to the simple parametrization of
approximative neutrino heating and cooling terms.
In self-consistent simulations with
neutrino transport the trend of easier 2D explosions with
higher angular resolution also holds for models with 
enhanced radial resolution~\cite{Summa2016}.

In contrast to the Garching code, schemes used by other 
authors (mostly with cartesian grids)
create perturbations for numerical reasons. Runs with
higher grid resolution may possess a lower level of such 
intrinsic noise and correspondingly show less favorable
conditions for the growth of hydrodynamic instabilities,
delaying the explosion. A verification of this possibility
will require detailed comparisons of simulations for well
defined test-setups.

\subsubsection{Assessment}

Some of the conflicting findings reported in
the literature may be a consequence of different codes
and setups with different levels of numerical noise and different
grid resolution in different domains of the simulations.
Directly comparing angular resolution of higher-order solvers
on polar grids with ``effective'' angle resolution of
cartesian grids~\cite{Couch2015} is extremely misleading.
Some of the seemingly contradictive conclusions may also be
traced back to the use of different parametric modeling approaches
with different simplifications of complex microphysics and
different quantities varied or kept fixed. Finally, also
the lack of self-consistency and the omission of feedback
mechanisms that couple hydrodynamics and neutrino emission
may be the cause of misleading results. Cautious interpretation
is therefore advisable and far-reaching conclusions should be
avoided.

\subsection{Turbulence vs.\ Convection vs.\ SASI: an Ongoing Debate}

The trend of later explosions in 3D compared to 2D
\cite[superimposed by considerable case-to-case stochasticity,
see][]{Takiwaki2014} is understood as a consequence
of the turbulent cascading of energy to small scales in 3D,
whereas in 2D energy is pumped into the largest 
structures~\cite{Hanke2012,Couch2013,Couch2014,Dolence2013}.

\subsubsection{Turbulent Kinetic Energy Spectrum in 2D and 3D}

\textbf{Figure~\ref{fig:turbulence}} shows energy spectra
$E(l)$ of nonradial (longitudinal) mass motions
based on a decomposition of the azimuthal velocity $v_{\theta}$
at a given radius (located in the gain layer) into spherical
harmonics $Y_{lm}(\theta,\phi)$:
\begin{equation}
E(l) = \sum^{l}_{m=-l}\left|\int_{\Omega}Y_{lm}^{*}(\theta,\phi) \,
\sqrt{\rho} \, v_{\theta}(r,\theta,\phi)\,\mathrm{d}\Omega\right|^{2}
\label{eq:turbpower}
\end{equation}
[cf.~Ref.~\cite{Hanke2012} for details; for a discussion of
differences between spectra for full 3D velocities and velocity
components, see Ref.~\cite{Radice2015}]. As expected
from the direction of the energy cascade, in 2D the lowest
modes clearly dominate the energy spectrum and
the spectral decline towards high spherical
harmonics modes (small wavelengths) is steeper (roughly $\propto
l^{-3}$) in the inertial range of the forward enstrophy cascade.
In contrast, in 3D the spectral decay in the inertial range of
the forward energy cascade is closer to $\propto l^{-5/3}$.
While this basic 
structure of the energy spectra was confirmed by many groups
\cite[e.g.,][]{Couch2013,Couch2014,Dolence2013,Abdikamalov2015,Takiwaki2014,Handy2014},
vivid twitter started about the exact shape of the power-law decline.
Some authors interpreted their 3D results by an $l^{-1}$
decay, piecewise power laws, or an exponential slope instead of 
the $l^{-5/3}$ cascade expected for
Kolmogorov turbulence~\cite{Kolmogorov1941,Landau1959} and 
surmised that anisotropic turbulence 
\cite[the radial component of the Reynolds stress is in rough
equipartition with the summed tangential components;][]{Murphy2013,MurphyMeakin2011},
non-stationarity or flow compressibility
might be responsible for non-Kolmogorov behavior. However,
Radice et al.~\cite{Radice2015,Radice2016} showed that increasing 
resolution enables the development of a power law with exponent
$(-5/3)$ in an increasing inertial range of wave numbers.

\subsubsection{Big Bubbles or Turbulence as Explosion Drivers}

The energy cascade in 2D thus feeds the biggest buoyant 
plumes, which have 
been recognized as conducive for driving shock
expansion~\cite{Hanke2012} because of their better
volume-to-surface ratio, enabling buoyancy to win over
drag effects~\cite{Couch2013,Fernandez2014}. 
3D is less favorable 
for the creation of such big bubbles through convective and 
turbulent processes, for which reason Hanke et al.\ (2012)
emphasized the possible importance of the SASI as a natural
mechanism to foster the growth of low-mode asymmetries
and to push shock expansion, provided the saturation amplitude
of SASI modes (which is limited by parasitic effects like
Rayleigh-Taylor and Kelvin-Helmholtz instabilities) is
sufficiently large.

Others continued to favor buoyancy-driven 
convection~\cite{Burrows2012,Murphy2013,Ott2013},
``penetrative convection''~\cite{Handy2014},
and ``turbulent convection''~\cite{MurphyMeakin2011,Abdikamalov2015}
as the main supportive instability
in the transition from the stalled shock to outward shock
acceleration. Couch \& Ott~\cite{Couch2015}, inspired 
by Murphy et al.~\cite{Murphy2013},
proposed that turbulence provides 
effective pressure that adds to the gas pressure in stabilizing
the gain layer, thus favoring the accumulation of energy, mass,
and momentum behind the shock and lowering the critical
neutrino luminosity for explosion. This hypothesis was motivated
by their observation that multidimensional simulations explode 
with lower neutrino-heating rates than 1D simulations. 
In this scenario enhanced neutrino heating by the convectively
stretched residence time of matter in the gain layer and 
turbulent Reynolds stresses conspire in fostering shock revival.
Couch and Ott~\cite{Couch2015} conclude a direct correlation between
the strength of turbulence and the susceptibility to explosion.

\subsubsection{Artifacts of Modeling}

The setups investigated by Radice et al.~\cite{Radice2016},
Burrows et al.~\cite{Burrows2012} and 
Dolence et al.~\cite{Dolence2013}
on the one hand and those of Couch \& Ott~\cite{CouchOtt2013,Couch2015}
on the other hand are probably not representative for 
supernovae and their progenitors in general. In the former 
works the shock stagnation radius is very big (exceeding
300\,km) and lingers there in a quasi-steady state for many
100\,ms. This large, quasi-stationary shock radius is likely to
be an artifact of the parametrized nuclear photodisintegration
and neutrino treatment. It is not typical of the shock behavior
in more realistic simulations with proper nuclear and neutrino
physics, where the shock initially expands to only
$\sim$150\,km, then tends to retreat again, and finally
accelerates outwards quickly once it achieves to expand beyond
the nucleon dissociation/recombination radius of 
$R_\mathrm{diss}\sim G M_\mathrm{NS}m_\mathrm{B}
/(8.8\,\mathrm{MeV})\sim 240\,(M_\mathrm{NS}/1.5\,M_\odot)$\,km
($m_\mathrm{B}$ is the baryon mass).

In the works of Couch \& Ott~\cite{CouchOtt2013,Couch2015},
large-amplitude, large-scale velocity 
perturbations in the Si/O layer of the progenitor model were
assumed. But even in their simulations without such an
imposed velocity field, large 
nonradial mass motions develop in the gain layer already 
20--30\,ms after core bounce, in contrast to the simulations
by the Garching group, where significant nonradial flows in 
the gain layer start only $\gtrsim$100\,ms post bounce.
The rapid growth of turbulence points to a sizable amplitude 
of numerical, probably grid-induced, perturbations in the
models of Refs.~\cite{CouchOtt2013,Couch2015}, which could 
instigate buoyancy in an uncontrolled way and could thus 
lead to an overestimated importance of turbulent convection.
The implications of such purely numerical effects for modeling
supernovae must still be clarified.

\subsubsection{Relevance of the SASI}
\label{sec:sasirelevance}

SASI-dominated phases were observed in self-consistent
3D supernova simulations 
\cite[e.g.,][]{Hanke2013,Tamborra2013,Tamborra2014a,Tamborra2014b,Melson2015b} 
as well as parametric studies \cite{Couch2014,Abdikamalov2015,Fernandez2015}
but were weak
in others, e.g.\ Ref.~\cite{Lentz2015} for a full-scale 
supernova model and 
Refs.~\cite{Ott2013,Burrows2012,Dolence2013,Couch2013,Couch2015}
for parametric models. Obviously,
unless SASI oscillations were damped by a lack of 
resolution \cite[cf.][]{Abdikamalov2015} or 
disfavored by the parameters of the modeling setup, the simulated
models seem to have passed through different dynamical regimes: 
the growth of the SASI is favored for small shock radii, 
in which case the SASI cycle time
is short (Eq.~\ref{eq:cycletime}) and its growth
rate large (Eq.~\ref{eq:omegasasi}), whereas convective activity
is facilitated by bigger shock radii (Eq.~\ref{eq:chiparameter}) 
or by greater numerical or imposed perturbations in the infalling
stellar core matter (Eq.~\ref{eq:nonlinear}).

Burrows et al.~\cite{Burrows2012}, tentatively seconded by
Couch \& O'Connor~\cite{Couch2014}, hypothesized that neutrino-driven
buoyant convection should almost always dominate the SASI in
neutrino-powered supernova explosions. However,
M\"uller et al.~\cite{Mueller2012sasi} and
Fernandez et al.~\cite{Fernandez2014,Fernandez2015} 
demonstrated that neutrino-driven explosions can develop
from both the regimes where SASI or convection dominate the
dynamics. SASI mass motions, in particular spiral modes, 
act as a storage of kinetic energy on large scales and
create secondary shocks that heat the matter by dissipating
kinetic energy. Both effects aid shock expansion.

\subsubsection{Assessment}

Despite strong opinions in favor of exclusive explanations,
a more likely possibility is that all effects of nonradial
flows ---convective buoyancy, turbulent pressure, and
SASI motions--- can influence the supernova shock dynamics
and could provide support to the postshock layer.
The relative importance of different nonradial instabilities
can depend on the phase of the post-bounce evolution and on
the detailed expansion and contraction behavior of the stalled
shock, which is sensitive to progenitor-specific properties of
the accretion flow. Present results of simulations may also be
affected by artifacts connected to the numerical grid and by
technical features in the (simplified) modeling setups.
Future, well-resolved and fully self-consistent 3D
simulations for larger sets of progenitors and realistic
pre-collapse perturbations in codes with low intrinsic noise
level are needed to confirm our expectation that the cores of 
collapsing stars can evolve through SASI-dominated episodes at
least transiently.

\begin{figure}[!]
\includegraphics[width=15cm]{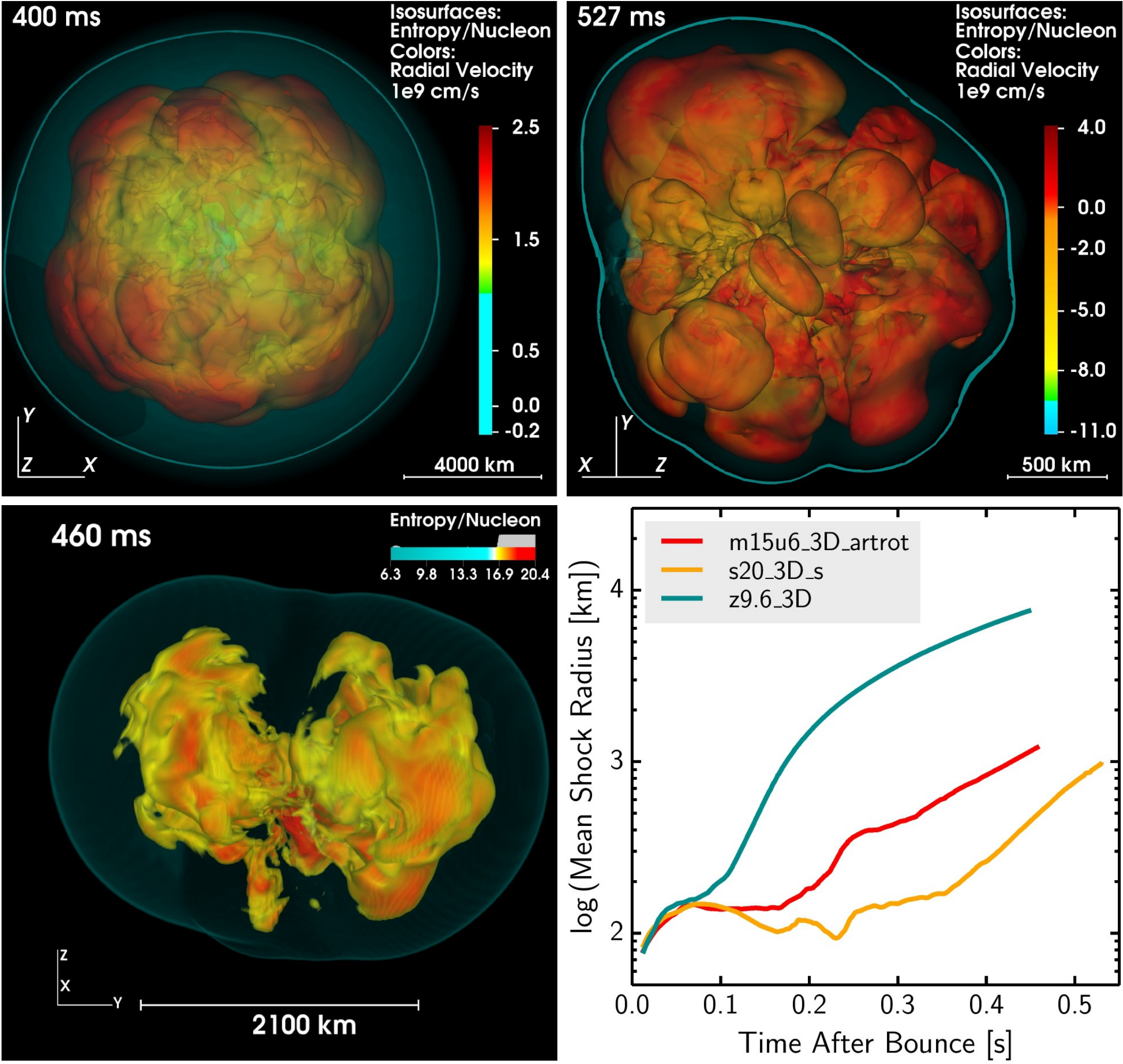}
\caption{Successful 3D explosion models of the Garching group
obtained in self-consistent neutrino-hydrodynamics simulations
with the \textsc{Prometheus-Vertex} code. The panels show 
isoentropy surfaces of neutrino-heated, buoyant matter for
a 9.6\,$M_\odot$ star \cite[top left;][]{Melson2015a},
a 20\,$M_\odot$ progenitor \cite[top right;][]{Melson2015b},
and a rotating 15\,$M_\odot$ model 
\cite[bottom left;][]{Summa2016b}.
The supernova shock is visible as a blue,
enveloping surface. The average shock radii as functions of
time are displayed in the lower right panel.}
\label{fig:3dexpl}
\end{figure}

\subsection{Self-consistent 3D Explosion Models}
\label{sec:3Dexplosions}

Successful neutrino-driven explosions were recently
obtained in self-consistent 3D simulations by several 
groups using different neutrino-transport approximations,
all based on RbR or RbR+ treatments of the angle
dependence on the employed coordinate grids.

\subsubsection{Recent Results}

Takiwaki et al.~\cite{Takiwaki2012,Takiwaki2014}, 
applying an IDSA 
code for $\nu_e$ and $\bar\nu_e$ transport in combination
with a leakage scheme for $\nu_x$, reported explosions 
for an 11.2\,$M_\odot$ star, but their models had modest 
resolution of $\sim$300 nonequidistant radial zones and 
at best 2.8$^\circ$ in the polar and azimuthal directions.
M\"uller~\cite{Mueller20152d3d} also obtained an explosion of this
progenitor, using a stationary two-stream solution of the RbR
transport equation combined with an analytic variable Eddington
factor closure~\cite{Mueller2015} and higher 
resolution (550$\times$128$\times$256 zones in $r$,
$\theta$, and $\phi$ directions).

The Oak Ridge and Garching groups found explosions
with more sophisticated, energy-dependent, three-flavor
transport solvers including more microphysics and 
intermediate numerical resolution. Oak Ridge, employing the
\textsc{Chimera} code with flux-limited diffusion and
540 logarithmically spaced radial zones and a
180$\times$180 zone (non-uniform)
$\theta$-$\phi$ grid, investigated a
15$\,M_\odot$ progenitor~\cite{Lentz2015}.
At Garching, explosions were obtained for 
9.6\,$M_\odot$ and 20\,$M_\odot$
stars~\cite{Melson2015a,Melson2015b} and for a rotating
15\,$M_\odot$ progenitor
\cite[\textbf{Figure~\ref{fig:3dexpl}};][]{Summa2016b}
with the \textsc{Prometheus-Vertex}
code, using a two-moment transport scheme with
Boltzmann closure, 400--600 nonequidistant radial
zones (improving the resolution dynamically) and 2$^\circ$
cell size in both angular directions of 
axis-free Yin-Yang~\cite{WongwathanaratHammer2010}
and standard polar grids. 

\subsubsection{New Messages}

The results of all these simulations agree in their basic
finding that 3D models are less susceptible to explosion
than 2D models. Shock revival, if happening at all, occurs
later in 3D than in 2D.
This outcome is in line with conclusions drawn from
most parametric studies (cf.~Sect.~\ref{sec:nonradflow}). 
None of the mentioned models could so far be
evolved to the point where the final explosion energy
was determined.
In earlier simulations of 11.2, 20, and 27\,$M_\odot$
stars the Garching team had not seen successful shock 
revival until $>$500\,ms after bounce, although the 
corresponding 2D models had already developed explosions
at that time 
\cite{Hanke2013,Tamborra2014a,Tamborra2014b,Tamborra2013}. 
Diagnostic parameters
that signal the proximity to explosion like the 
maximum shock radius and the ratio of advection to heating
time scale suggested marginal failures. Indeed, in the
recent 20\,$M_\odot$ simulation~\cite{Melson2015b}
a reduction of neutral-current
neutrino-nucleon scattering by effectively 10--20\%
in the neutrinospheric layers (motivated by possible 
strangeness contributions to the nucleon spin, affecting
the axial-vector weak coupling)
led to an increase of the neutrino
luminosities and of the neutrino heating behind the 
shock that was sufficient to turn the failed explosion
to success. Although for the purpose of demonstration
the magnitude of the considered strangeness correction
was overestimated compared to the currently best 
experimental and theoretical limits~\cite{Hobbs2016},
the result of Melson et al.~\cite{Melson2015b}
calls attention to an important sensitivity of
3D supernova models to even only smaller variations of the 
neutrino opacities.

Interestingly, the 9.6\,$M_\odot$ explosion of
Melson et al.~\cite{Melson2015a} developed a slightly higher
explosion energy than the corresponding 2D case. This
can be explained by differences of the dynamics of
convective downflows. In 3D Kelvin-Helmholtz instability
leads to more efficient fragmentation, which decelerates
the downdrafts and keeps more matter in the gain layer,
thus reducing neutrino-energy losses below the gain
radius. M\"uller~\cite{Mueller20152d3d} observed a similar,
even larger effect in his successful 11.2\,$M_\odot$ model,
which exploded earlier and more powerfully in 3D
than in 2D 
\cite[in contrast to the results by][]{Takiwaki2012,Takiwaki2014}.
He traced
the more favorable 3D situation back to a variety of
subtle differences in the accretion and outflow
dynamics in 2D and 3D, leading to more efficient
driving of neutrino-heated gas outflow in 3D.
Whether these interesting 3D effects also apply to
more massive progenitors and how they might depend
on numerical resolution still needs to be figured out.

\subsubsection{Implications}

These results suggest that the discussion of 3D 
effects in comparison to 2D involves at least two 
separate aspects:
\begin{enumerate}
\item What is the dimensionality dependence of the
dynamics prior to the onset of explosion? Do 3D 
simulations develop shock revival faster or more
delayed than in 2D? The majority of current models
(with few exceptions) suggests the latter.
\item What are 3D vs.\ 2D differences after the onset
of the explosion? Can 3D flow dynamics enhance the 
explosion energy and thus bring 3D models closer to
observed supernova energies? Here the results of 
Refs.~\cite{Melson2015b,Mueller20152d3d}
suggest interesting advantages of the 3D case.
Parametric explosion studies in Ref.~\cite{Handy2014}
seem to yield support.
\end{enumerate}

Overall, one is therefore tempted to conclude that the 
self-consistent 3D calculations provide back-up for the
neutrino-driven mechanism. When the models fail, not much
seems to be missing to achieve shock revival. The 2D and
3D simulations of the Oak Ridge group (also those of
Series~C) are considerably more optimistic than those 
from Garching, despite similarities in many aspects of
the numerical modeling 
\cite[but also differences in quite a few, cf.][]{Lentz2015}.
The reason for this 
discrepancy is unclear and can be clarified only by
detailed and direct comparisons.

\subsubsection{Caveats}

A serious drawback of the current self-consistent 
simulations are the constraints of numerical resolution
because of the enormous computational demands when
detailed microphysics and transport are included.
Radice et al.~\cite{Radice2015,Radice2016}, 
assuming that steady-state
turbulence applies and setting up suitable conditions, 
demonstrated that Kolmogorov scaling can progressively
be recovered as the resolution in their toy models
is increased. Although they found that the resolution 
of the published state-of-the-art supernova simulations
seems to be sufficient to describe gross features
of neutrino-driven turbulent convection, the results of
Refs.~\cite{Radice2015,Radice2016} for successive grid refinements
exhibit subtle and partly non-monotonic differences, whose
relevance for supernova dynamics is currently unclear. Better
resolution in particular diminishes a ``bottleneck effect'',
in which numerical viscosity hinders the efficient cascading
of turbulent energy to small scales and keeps energy on large
scales. Abdikamalov et al.~\cite{Abdikamalov2015}
expressed concerns that
this bottleneck might incorrectly and artificially promote
explosions. It will have to be tested whether present
self-consistent supernova simulations are affected by such
artifacts as soon as higher-resolution calculations can be
afforded.

\subsection{A Universal Explosion Criterion}

Simulations at the onset of runaway shock expansion exhibit a wide
range of variation of their diagnostic parameters, e.g.\ of their neutrino 
luminosities, average and maximum shock radii, mass accretion rate, 
total mass in the gain layer and mass fraction of recombined nucleons
there, neutrino-heating rate and efficiency, average and maximum
entropy, turbulent kinetic energy in the gain layer, etc. The 
question therefore arises whether there is any individual parameter
value or parameter relation that has to be matched when shock revival 
shall occur? Could such a criterion capture the influence
of dimensionality and of multidimensional effects on the 
susceptibility to explosion?

Burrows \& Goshy~\cite{BurrowsGoshy1993} proposed a critical luminosity 
condition as a function of the mass-accretion rate of the 
stalled shock, $L_{\nu,\mathrm{crit}}(\dot M)$, which defines  
the threshhold value of the neutrino luminosity that needs to be
exceeded to enable shock expansion against the ram pressure of the
infalling material. Their semi-analytic analysis and 
follow-up works (numerical and analytical) by others
confirmed the existence of an upper limit of the neutrino-energy
deposition in the gain layer that is compatible with solutions for
quasi-stationary accretion shocks 
\cite[for a review, see][]{Janka2012}.

Although there are interesting proposals of alternative explosion
criteria like an antesonic condition~\cite{Pejcha2012,MurphyMeakin2011} 
and an integral condition for the gain layer~\cite{Murphy2015},
the considerations here will focus on a generalization of the critical
luminosity relation.

\subsubsection{Critical Luminosity in Spherical Symmetry}

Janka~\cite{Janka2012} argued 
\cite[see also][]{Mueller2015} that the
radius of the stalled shock in 1D models approximately follows
the scaling relation 
\begin{equation}
R_\mathrm{s}\propto\frac{\left(L_\nu\langle E_\nu^2\rangle\right)^{4/9}
R_\mathrm{g}^{16/9}}{\dot{M}^{2/3}M_\mathrm{NS}^{1/3}}\,,
\label{eq:rshock1}
\end{equation}
where $L_\nu$ is defined as the total
luminosity of $\nu_\mathrm{e}$ plus $\bar{\nu}_\mathrm{e}$,
$L_\nu=L_{\nu_\mathrm{e}}+L_{\bar{\nu}_\mathrm{e}}$,
and $\langle E_\nu^2\rangle$ denotes the weighted average of the
corresponding mean squared energies:
\begin{equation}
\langle E_\nu^2\rangle = \frac{L_{\nu_\mathrm{e}} 
\langle E_{\nu_\mathrm{e}}^2\rangle + L_{\bar{\nu}_\mathrm{e}} 
\langle E_{\bar{\nu}_\mathrm{e}}^2\rangle}{L_{\nu_\mathrm{e}}+L_{\bar{\nu}_\mathrm{e}}}\,.
\end{equation}
Both determine the neutrino-energy deposition in the gain layer,
which is given by
\begin{equation}
\dot{Q}_\nu\propto\frac{L_{\nu_\mathrm{e}}\langle
E_{\nu_\mathrm{e}}^2\rangle+L_{\bar{\nu}_\mathrm{e}}\langle
E_{\bar{\nu}_\mathrm{e}}^2\rangle}{R^2_\mathrm{g}}\,M_\mathrm{g}\,,
\label{eq:qheat}
\end{equation}
where $\langle E_{\nu_i}^2\rangle \equiv 
\langle\epsilon_{\nu_i}^3\rangle/\langle\epsilon_{\nu_i}\rangle$
(for $\nu_i = \nu_e,\,\bar\nu_e$)
are defined as squared energies of the neutrino-energy distributions
expressed in terms of the energy moments of the neutrino-number
distributions.

Janka~\cite{Janka2012} showed that the critical luminosity condition 
can be deduced from equating the advection time scale,
\begin{equation}
\tau_\mathrm{adv} \approx 
\frac{R_\mathrm{s}-R_\mathrm{g}}{\beta^{-1}|v_\mathrm{ff}(R_\mathrm{s})|}
\propto \frac{R_\mathrm{s}^{3/2}}{\sqrt{M_\mathrm{NS}}}\,,
\label{eq:tadv}
\end{equation}
with the heating time scale,
\begin{equation}
\tau_\mathrm{heat} = \frac{E_\mathrm{tot,g}}{\dot{Q}_\nu} \propto 
\frac{\left|\bar{e}_\mathrm{tot,g}\right|R_\mathrm{g}^2}{L_\nu\langle E_\nu^2\rangle} \,.
\label{eq:theat}
\end{equation}
Here, $\bar{e}_\mathrm{tot,g}$ is the average
mass-specific binding energy in the gain layer,
\begin{equation}
\bar{e}_\mathrm{tot,g} = \frac{E_\mathrm{tot,g}}{M_\mathrm{g}}\,,
\end{equation}
where $E_\mathrm{tot,g}$ is the total (internal plus gravitational plus
kinetic energy) and $M_\mathrm{g}$ the mass in the gain layer. 
In Eq.~(\ref{eq:tadv}) the approximations are used that 
$M_\mathrm{NS}\gg M_\mathrm{g}$ and, roughly, $R_\mathrm{s}\gg R_\mathrm{g}$.
Employing $\tau_\mathrm{adv}/\tau_\mathrm{heat} = 1$ as well established 
condition that signals runaway shock expansion 
\cite[for a detailed discussion, see, e.g.,][]{Fernandez2012}
one derives \cite[cf.\ also][]{Mueller2015}:
\begin{equation}
\left(L_\nu \langle E_\nu^2\rangle\right)_\mathrm{crit}\propto
\left(\dot{M}M_\mathrm{NS}\right)^{3/5}\left|\bar{e}_\mathrm{tot,g}\right|^{3/5}R_\mathrm{g}^{-2/5}.
\label{eq:lcrit1}
\end{equation}
Following Ref.~\cite{Summa2016} we do not omit $\bar{e}_\mathrm{tot,g}$ 
and $R_\mathrm{g}$ in this relation.

The path to generalize this condition to the multidimensional case
was exemplified by M\"uller \& Janka (2015) and Summa et al.\ (2015).
Despite the complex nature of the postshock flow including
violent, large-amplitude SASI sloshing, bubble buoyancy, and 
turbulent convection and flow fragmentation
in a highly non-stationary environment, these works found that the
overall effects of nonradial mass motions seem to be captured 
astonishingly well by a simple concept described in the following.

\subsubsection{Effects of Turbulence}
\label{sec:turbeffects}

Guided by Ref.~\cite{Mueller2015} we take multidimensional
(``turbulent'') mass motions in the gain layer into account 
through an isotropic pressure contribution that is coined in terms
of the squared Mach number of unordered fluid flows, averaged over the
gain region: $P_\mathrm{turb}\approx
\langle v_\mathrm{aniso}^2\rangle\rho\approx4/3\langle\mathrm{Ma}^2\rangle P$,
ignoring additional complexity, e.g., by turbulent energy transport
or centrifugal support \cite[cf.][]{Radice2015,MurphyMeakin2011}.
Including this additional postshock pressure in the shock-jump condition,
one finds a larger radius of the stalled shock compared to
Eq.~(\ref{eq:rshock1}):
\begin{equation}
R_\mathrm{s}\propto\frac{\left(L_\nu\langle E_\nu^2\rangle\right)^{4/9}
R_\mathrm{g}^{16/9}}{\dot{M}^{2/3}M_\mathrm{NS}^{1/3}}\,\xi_\mathrm{turb}^{2/3}
\label{eq:rshock2}
\end{equation}
with 
\begin{equation}
\xi_\mathrm{turb} \equiv 1 + \frac{4}{3}\langle\mathrm{Ma}^2\rangle \,.
\label{eq:xiturb}
\end{equation}
Through a corresponding stretching of the advection time scale 
(Eq.~\ref{eq:tadv}), the condition $\tau_\mathrm{adv} =
\tau_\mathrm{heat}$ leads to a modified critical 
luminosity~\cite{Mueller2015,Summa2016}: 
\begin{equation}
\left(L_\nu \langle E_\nu^2\rangle\right)_\mathrm{crit}\propto
\left(\dot{M}M_\mathrm{NS}\right)^{3/5}\left|\bar{e}_\mathrm{tot,g}\right|^{3/5}
R_\mathrm{g}^{-2/5}\,\xi_\mathrm{turb}^{-3/5} \,.
\label{eq:lcrit2}
\end{equation}

Different from Refs.~\cite{Mueller2015,Summa2016},
we empirically choose the total anisotropic kinetic energy, 
$E_\mathrm{kin,g}^\mathrm{aniso}$, in the definition of 
$\langle\mathrm{Ma}^2\rangle$ instead of just the lateral component:
\begin{equation}
\langle\mathrm{Ma}^2\rangle = 
\frac{\langle v_\mathrm{aniso}^2 \rangle}{\langle c_\mathrm{s,g}^2\rangle} =
\frac{\langle(v_r-\bar{v}_r)^2\rangle+\langle(v_\theta-\bar{v}_\theta)^2\rangle
+\langle(v_\phi-\bar{v}_\phi)^2\rangle}{\langle c_\mathrm{s,g}^2\rangle} =
\frac{2E_\mathrm{kin,g}^\mathrm{aniso}/M_\mathrm{g}}{\langle c_\mathrm{s,g}^2\rangle}\,,
\label{eq:mach}
\end{equation}
with $\bar{v}_{r,\theta,\phi}$ being angular averages over
spherical shells. This generalization shall make Eq.~(\ref{eq:lcrit2})
applicable to 2D and 3D results with the
same proportionality factor. The velocity
differences in Eq.~(\ref{eq:mach}) are meant to subtract ordered 
radial flows like accretion or expansion of the gain layer and coherent
angular motions associated with stellar rotation and spiral SASI modes.
In contrast to M\"uller \& Janka~\cite{Mueller2015}, we do not apply an 
approximation for the sound speed behind the shock but extract 
$\langle c_\mathrm{s,g}^2\rangle$ directly from the numerical
simulations as mass-weighted average over the gain layer:
\begin{equation}
\langle c_\mathrm{s,g}^2 \rangle = 
\frac{1}{M_\mathrm{g}}\int\limits_{V_\mathrm{g}} 
\mathrm{d}V\,c_\mathrm{s}^2 \rho\,.
\end{equation}

For marginal stability of the gain layer to convection in 
a steady-state situation, the volume-integrated neutrino heating
rate must be balanced by outward ``turbulent luminosity'', i.e.\
$\dot Q_\nu \sim L_\mathrm{turb}$~\cite{Murphy2013}.
According to M\"uller \& Janka~\cite{Mueller2015}, this requirement 
is reflected by the following relation between
$E_\mathrm{kin,g}^\mathrm{aniso}$ and $\dot Q_\nu$:
\begin{equation}
\frac{E_\mathrm{kin,g}^\mathrm{aniso}}{M_\mathrm{g}}\propto 
\left[\left(R_\mathrm{s}-R_\mathrm{g}\right)
\frac{\dot{Q}_\nu}{M_\mathrm{g}}\right]^{2/3}.
\label{eq:eaniso}
\end{equation}
Here $R_\mathrm{s}$ and $R_\mathrm{g}$ denote 
angle averages of the shock and gain radius, respectively.
This relation turns out to be well fulfilled when the average
shock radius is nearly stationary, but large deviations occur
when the shock moves.

\begin{figure}[!]
\includegraphics[width=9.0cm]{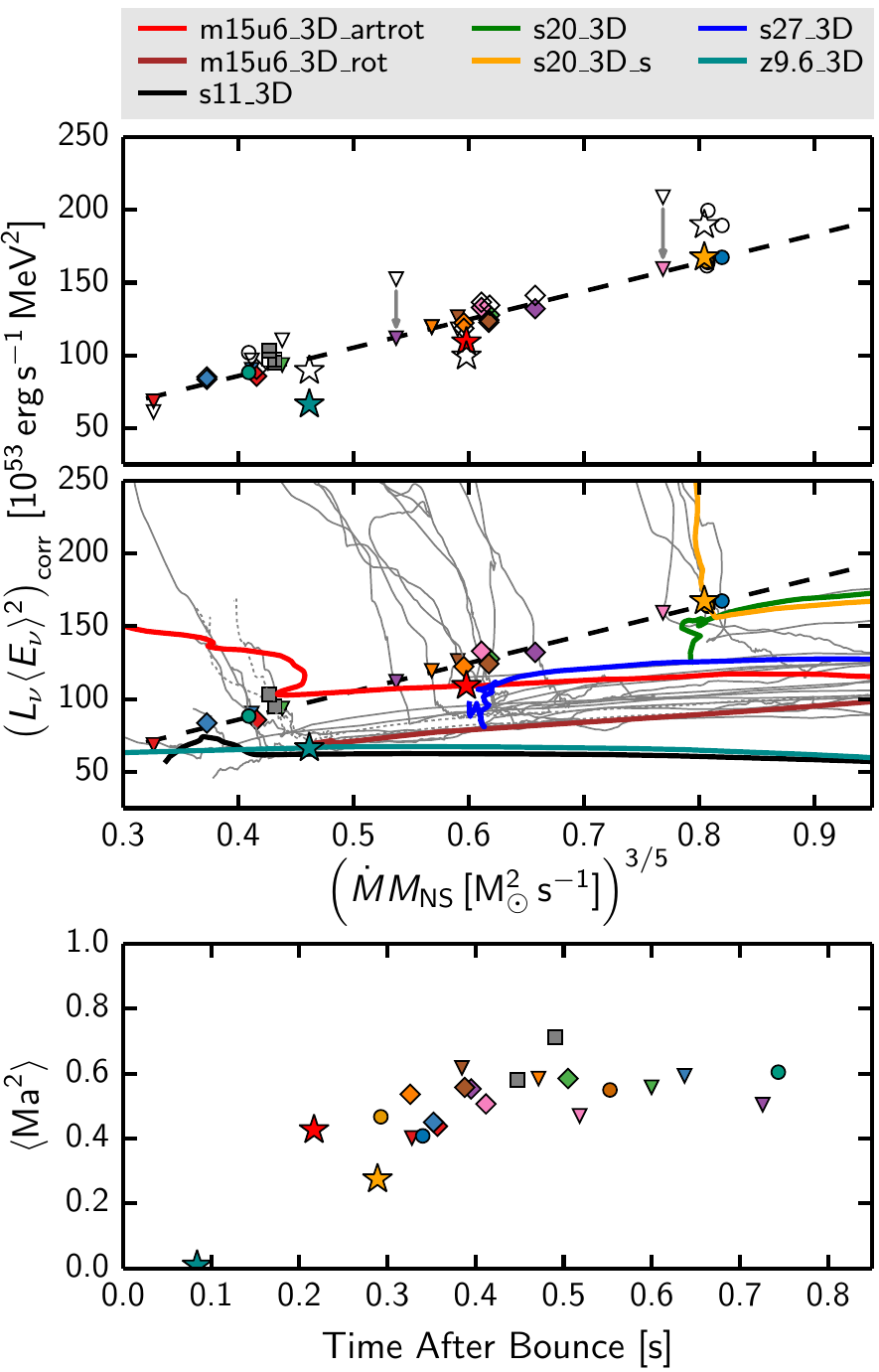}
\caption{Critical luminosity condition for explosion.
The upper two panels depict the critical relation between
$\left(L_\nu\langle E_\nu^2\rangle\right)_\mathrm{crit,corr}$
and $\left(\dot{M}M_\mathrm{NS}\right)^{3/5}$ for the onset of
explosion (Eq.~\ref{eq:critcorr}) as black dashed line obtained
from a least-squares fit to the critical points of the 
2D model set (triangles, circles and diamonds) of
Ref.~\cite{Summa2016}. In addition, results of rotating 2D 
(squares) and 3D models (stars) are displayed. Open
symbols show the locations of the uncorrected values of
$\left(L_\nu\langle E_\nu^2\rangle\right)_\mathrm{crit}$
(Eq.~\ref{eq:lcrit4}), arrows indicate shifts by the
correction factor $(\xi_\mathrm{g}/\xi_\mathrm{g}^*)^{-1}$
of Eq.~(\ref{eq:critcorr}) in some cases.
In the middle panel the evolution tracks (from
right to left) of exploding and 
nonexploding 2D (gray), rotating 2D (gray, dashed)
and 3D models (colored, legend on top; for exploding
cases see \textbf{Figure~\ref{fig:3dexpl}}) are drawn. Exploding
models cross the critical line.
The bottom panel depicts average squared ``turbulent'' Mach numbers 
in the gain layer (Eq.~\ref{eq:mach}) at the time when runaway 
shock expansion begins. The onset
of the explosion in model z9.6$\_$3D does not coincide
with the critical line because its initiation by buoyant plumes is not
compatible with the conceptual framework for deriving the luminosity
correction. 
All plotted values were smoothed with running averages of 25\,ms.
Due to storage constraints of 3D data $\langle E_\nu\rangle^2$
was used instead of $\langle E_\nu^2\rangle$.
}
\label{fig:corr}
\end{figure}

\subsubsection{Effects of Rotation}
\label{sec:roteffects}

The effects of rotation can be included in a similar way. Rotation
provides centrifugal support in the infall region ahead 
of the shock. Instead of the free-fall velocity of the matter,
$v_\mathrm{ff}(R_\mathrm{s}) = -\sqrt{2GM_\mathrm{NS}/R_\mathrm{s}}$, 
which was used in Eq.~(\ref{eq:tadv}), one now gets
\begin{equation}
v_\mathrm{coll}(R_\mathrm{s}) = - \sqrt{v_\mathrm{ff}^2(R_\mathrm{s})
-\frac{j_0^2}{R_\mathrm{s}^2}} = \xi_\mathrm{rot}\,v_\mathrm{ff}(R_\mathrm{s})
\label{eq:vcoll}
\end{equation}
from solving the equation of motion, $\mathrm{d}v_r/\mathrm{d}t 
= -GM_\mathrm{NS}/r^2 + j_0^2/r^3$, assuming conservation of the specific
angular momentum, $j_0$, during collapse. In Eq.~(\ref{eq:vcoll}) we have
introduced the rotational correction factor
\begin{equation}
\xi_\mathrm{rot} \equiv \sqrt{1 - \frac{j_0^2}{2GM_\mathrm{NS}R_\mathrm{s}}} \le 1\,.
\label{eq:xirot}
\end{equation}
Direction-averaging the effect of rotation, one can define $j_0$ as 
average of the specific angular momentum on spherical shells.
Considering in the postshock layer a modest radial increase of the 
spherically averaged specific angular momentum, 
$j = j_\mathrm{s}(r/R_\mathrm{s})^{1/2}$~\cite{Burasetal2006B},
centrifugal effects in rotational equilibrium can be absorbed into a
correction of the gravitating mass, $M'_\mathrm{NS} = 
M_\mathrm{NS}\,[1-j_\mathrm{s}^2/(GM_\mathrm{NS}R_\mathrm{s})]$.
The usual equation of hydrostatic equilibrium thus applies with 
$M_\mathrm{NS}$ being replaced by $M'_\mathrm{NS}$. Therefore the 
density, temperature, and pressure profiles in the relativistic 
gas-pressure dominated gain layer can still be approximated by
$\rho\propto r^{-3}$, $T\propto r^{-1}$, $P \propto T^4\propto
\rho T\propto r^{-4}$~\cite{Janka2001A}. Making use of this result and 
of Eq.~(\ref{eq:vcoll}) for the preshock velocity, a derivation in 
analogy to the one in Refs.~\cite{Janka2012,Mueller2015}
yields for the radius of the stalled shock in the presence of rotation:
\begin{equation}
R_\mathrm{s}\propto\frac{\left(L_\nu\langle E_\nu^2\rangle\right)^{4/9}
R_\mathrm{g}^{16/9}}{\dot{M}^{2/3}M_\mathrm{NS}^{1/3}}\,\xi_\mathrm{rot}^{-2/3}
\,.
\label{eq:rshock3}
\end{equation}
Accounting for both this change of the shock-stagnation radius and for
the rotational deceleration of the infall velocity in the postshock
advection time scale (Eq.~\ref{eq:tadv}), the condition $\tau_\mathrm{adv}
= \tau_\mathrm{heat}$ yields now:
\begin{equation}
\left(L_\nu \langle E_\nu^2\rangle\right)_\mathrm{crit}\propto
\left(\dot{M}M_\mathrm{NS}\right)^{3/5}\left|\bar{e}_\mathrm{tot,g}\right|^{3/5}
R_\mathrm{g}^{-2/5}\,\xi_\mathrm{rot}^{6/5} \,.
\label{eq:lcrit3}
\end{equation}
Rotation leads to a larger shock-stagnation radius
(Eq.~\ref{eq:rshock3} with $\xi_\mathrm{rot} < 1$). It also decreases the
critical luminosity because of the factor $\xi_\mathrm{rot}^{6/5}$. 
In addition, rotational energy in the gain layer provides a 
positive contribution to $\bar{e}_\mathrm{tot,g} < 0$, shifting 
$\bar{e}_\mathrm{tot,g}$ closer to zero, which also decreases the
rhs of Eq.~(\ref{eq:lcrit3}).

\subsubsection{Universal Critical Luminosity Condition}

Including effects of rotation as well as unordered (``turbulent'')
mass motions requires the combination of both factors $\xi_\mathrm{turb}$
and $\xi_\mathrm{rot}$ in the shock stagnation radius (cf.\ 
Eqs.~\ref{eq:rshock2} and \ref{eq:rshock3}), 
\begin{equation}
R_\mathrm{s}\propto\frac{\left(L_\nu\langle E_\nu^2\rangle\right)^{4/9}
R_\mathrm{g}^{16/9}}{\dot{M}^{2/3}M_\mathrm{NS}^{1/3}}\,
\left(\frac{\xi_\mathrm{turb}}{\xi_\mathrm{rot}}\right)^{\! 2/3} ,
\label{eq:rshock4}
\end{equation}
and in the critical luminosity criterion (cf.\ Eqs.~\ref{eq:lcrit2} and 
\ref{eq:lcrit3}),
\begin{equation}
\left(L_\nu \langle E_\nu^2\rangle\right)_\mathrm{crit}\propto
\left(\dot{M}M_\mathrm{NS}\right)^{3/5}\left|\bar{e}_\mathrm{tot,g}\right|^{3/5}
R_\mathrm{g}^{-2/5}\,\xi_\mathrm{turb}^{-3/5}\,\xi_\mathrm{rot}^{6/5}
\equiv \left(\dot{M}M_\mathrm{NS}\right)^{3/5}\xi_\mathrm{g}\,,
\label{eq:lcrit4}
\end{equation}
where the time-dependent quantity $\xi_\mathrm{g}$ subsumes all gain-layer 
related properties:
\begin{equation}
\xi_\mathrm{g}\equiv\left|\bar{e}_\mathrm{tot,g}\right|^{3/5}
R_\mathrm{g}^{-2/5}\,\xi_\mathrm{turb}^{-3/5}\,\xi_\mathrm{rot}^{6/5}\,.
\label{eq:xi}
\end{equation}
$\xi_\mathrm{g}$ can be used to correct $L_\nu \langle E_\nu^2\rangle$
for variations of the time evolution of gain radius,
binding energy, nonradial (turbulent) postshock flows, and rotation,
which lead to time- and model-dependent variations of the critical 
luminosity in addition to the basic dependence on $M_\mathrm{NS}$ and
$\dot{M}$. Doing so we derive a generalized, universal relation for
the critical luminosity~\cite[cf.][]{Summa2016}:
\begin{equation}
\left(L_\nu \langle E_\nu^2\rangle\right)_\mathrm{crit,corr}\equiv
\frac{1}{\xi_\mathrm{g}/\xi_\mathrm{g}^*}
\left(L_\nu \langle E_\nu^2\rangle\right)_\mathrm{crit}
\propto \left(\dot{M}M_\mathrm{NS}\right)^{3/5}\,.
\label{eq:critcorr}
\end{equation}
The arbitrary constant $\xi_\mathrm{g}^*$ is introduced as
normalization of the correction factor relative to a chosen reference
model, for which $\xi_\mathrm{g}^*$ is evaluated at the time when
$\tau_\mathrm{adv}/\tau_\mathrm{heat} = 1$.

\paragraph{Application to Self-consistent Supernova Models}
\textbf{Figure~\ref{fig:corr}} shows evolutionary tracks (running from
right to left) in the plane 
spanned by $\left(L_\nu \langle E_\nu^2\rangle\right)_\mathrm{corr} =
\left(L_\nu \langle E_\nu^2\rangle\right)/
(\xi_\mathrm{g}/\xi_\mathrm{g}^*)$ and
$\left(\dot{M}M_\mathrm{NS}\right)^{3/5}$ for the set of 2D models
of Ref.~\cite{Summa2016}. The three panels also include two
exploding 2D simulations with rotation 
as well as exploding and nonexploding 3D models
computed at Garching (see Sect.~\ref{sec:3Dexplosions}; the cases
with explosions are displayed in \textbf{Figure~\ref{fig:3dexpl}}). 
Colored symbols on top of the tracks mark the instants when the
explosion sets in (i.e.\ when $\tau_\mathrm{adv}/\tau_\mathrm{heat} = 1$)
and correspond to the critical luminosities of Eq.~(\ref{eq:critcorr}).
Open symbols in the top panel indicate the locations of the uncorrected
values of Eq.~(\ref{eq:lcrit4}). 

Obviously, for all 2D models the correction factor 
$(\xi_\mathrm{g}/\xi_\mathrm{g}^*)^{-1}$ successfully compensates for 
the influence of nonradial mass flows, rotation, and model-to-model
variations of gain radius and specific energy in the gain layer.
Consequently, the colored symbols obey a tight correlation as expected
from Eq.~(\ref{eq:critcorr}), and the dashed black line defines 
a universal critical explosion condition that holds for 2D simulations.

The remaining low-level scattering could be linked to
approximations that entered the derivation and evaluation of the 
scaling relation of Eq.~(\ref{eq:critcorr}), e.g.: simple power laws
for the radial structure of the gain layer; assumed scaling of
$\tau_\mathrm{adv}$ with postshock quantities; use of
$L_\nu$ and $\langle E_\nu^2\rangle$ as measured at infinity
instead of values at the gain radius; omission of gravity
contributions from the mass of the gain layer and pressure
corrections in computing the 
preshock velocity (Eq.~\ref{eq:vcoll});
simple averaging of rotation effects on spheres;
or the assumption of a model-independent
compression ratio $\beta$ in the shock.

The upward bending of the evolutionary tracks at the
onset of explosion is caused by a steep drop of $\xi_\mathrm{g}$
in the denominator
of $\left(L_\nu \langle E_\nu^2\rangle\right)_\mathrm{corr}$,
while $\left(L_\nu \langle E_\nu^2\rangle\right)$ in the numerator
evolves slowly. The decline of $\xi_\mathrm{g}$ (Eq.~\ref{eq:xi})
occurs because a strong increase of $\xi_\mathrm{turb}$ supports the 
outward acceleration of the shock and, as a consequence, the specific
binding energy of the gain layer 
($\left|\bar{e}_\mathrm{tot,g}\right|$) plummets.

\paragraph{Behavior of 3D Models}
The 3D models deserve special discussion because their 
behavior is less uniform. 

Except for model z9.6$\_$3D, all 3D simulations with successful
and failed explosions for nonrotating stars 
fit the general picture that applies
for the 2D cases. The evolution tracks of all nonexploding 3D
runs including the ones that marginally fail, remain below the 
(corrected) critical luminosity curve defined by the
2D simulations (see \textbf{Figure~\ref{fig:corr}}, upper two panels).
The nonrotating 3D model s20$\_$3D$\_$s~\cite{Melson2015b} 
explodes with about half the value of the average squared Mach number
($\langle\mathrm{Ma}^2\rangle \sim 0.28$; \textbf{Figure~\ref{fig:corr}},
bottom panel) of typical 2D models, but its critical point coincides
perfectly with the critical luminosity curve.

The low-mass model z9.6$\_$3D~\cite{Melson2015a} clearly
defines an outlier. Its explosion starts
($\tau_\mathrm{adv}/\tau_\mathrm{heat} = 1$) long before the
evolution track reaches the critical luminosity curve of
Eq.~(\ref{eq:critcorr}). Here the blast-wave acceleration
is driven by the outward expansion of buoyant plumes with
neutrino-heated matter, and postshock turbulence does not play 
an important role. Also the specific energy in the gain layer 
does not increase significantly when runaway shock expansion
occurs, because the rising plumes of dilute gas contain only
a minor fraction of the mass in the gain layer.
In fact, the {\em uncorrected} quantity
$\left(L_\nu \langle E_\nu^2\rangle\right)_\mathrm{crit}$
(Eq.~\ref{eq:lcrit4}) of model z9.6$\_$3D lies close to the
critical curve for all other models
(see open star in \textbf{Figure~\ref{fig:corr}}, top panel), 
and the corrected quantity is farther away from this curve.
This unfavorable trend is caused by the normalization
with the factor $\xi_\mathrm{g}^*$, which is picked from one
of the 2D explosion models and reflects the small 
$\xi_\mathrm{g}$-values of these cases.
This observation illustrates that
model z9.6$\_$3D blows up nearly like a 1D explosion. 
Under such circumstances normalization with $\xi_\mathrm{g}^*$ 
from multidimensional models is not appropriate.

The rapidly rotating model m15u6$\_$3D$\_$artrot also explodes,
in contrast to the more slowly spinning case m15u6$\_$3D$\_$rot,
which fails to blow up~\cite{Summa2016b}. At the onset of 
the explosion the value of $\langle\mathrm{Ma}^2\rangle$ in
model m15u6$\_$3D$\_$artrot is similar
to those of the 2D cases, although its evolutionary track indicates
the special, rotation-dominated post-bounce dynamics of this model
by passing the critical luminosity nearly horizontally before
bending upwards sharply (\textbf{Figure~\ref{fig:corr}}, middle panel).
Shock revival in this case is fostered by a strong spiral
SASI mode in agreement with findings in
Refs.~\cite{Nakamura2014,Iwakami2014}.
It is important to note that
preshock rotation plays a negligible role since
$\xi_\mathrm{rot} > 0.993$. The main effect of rotation is its
contribution of kinetic energy as part of the total energy in
the gain layer. In addition, spiral SASI waves also trigger
secondary turbulent mass motions, creating effective turbulent
pressure. Both effects are accounted for in our formalism
of Sects.~\ref{sec:turbeffects} and \ref{sec:roteffects}.
Sufficiently rapid rotation
can therefore facilitate the explosion and reduces the
critical threshold for the neutrino luminosity
(Eq.~\ref{eq:lcrit4}). This can be seen by the open star of
model m15u6$\_$3D$\_$artrot in the top panel of
\textbf{Figure~\ref{fig:corr}}. Although our correction
procedure of Eq.~(\ref{eq:critcorr}) moves the critical luminosity 
of this model closer towards the threshold line, it still lies
somewhat below this curve and agrees with the universal relation
less well than the other models. The crossing of the evolution
track of model m15u6$\_$3D$\_$artrot therefore
happens somewhat later than the equality 
$\tau_\mathrm{adv} = \tau_\mathrm{heat}$ is fulfilled. However,
since both the evolution track and the critical curve are fairly 
flat, small uncertainties in the vertical location can cause a
considerable shift of the crossing point. For this reason it is
not clear whether the slight mismatch of m15u6$\_$3D$\_$artrot
is a mere incidence or whether it points to shortcomings or 
incompleteness of the theoretical framework to describe the
effects of rotation in this case.

We conclude that the dynamical evolution of
model z9.6$\_$3D definitely and of m15u6$\_$3D$\_$artrot 
possibly is not well captured, or at
least not fully represented, by the theoretical concept that
underlies the derivation of Eq.~(\ref{eq:critcorr}).
The functional relation given by this equation seems to define
a universal critical luminosity condition for the subclass
of 2D and 3D models where strong, nonordered (``turbulent'') flows
play a major role for the postshock dynamics. Deviations from the
relation of Eq.~(\ref{eq:critcorr}) may signal cases 
where the approach to explosion 
is determined by alternative dynamical phenomena like 
bouyancy or spiral SASI motions that drive the shock expansion. 
A larger set of 3D models is needed, however, to consolidate
this framework.

\section{NEW PHENOMENA IN THREE DIMENSIONS}

The first self-consistent 3D simulations have already enabled
discoveries of new phenomena, which are associated with the
special dynamical behavior of 3D flows in the supernova core and
can cause characteristic imprints on the neutrino emission.
Predicting such signal features in detail is of crucial
importance for the interpretation of the neutrino measurement
from a future Galactic supernova.

Based on its particularly low photomultiplier dark noise rates,
IceCube is an excellent supernova burst detector by observing a
collective rise in all photomultiplier rates on top of the dark
noise. Although IceCube will not be able to provide directional
information and no event-by-event energy information, it is
unique among all existing SN neutrino observatories in its
ability to record the signal with a 2\,ms timing resolution.
IceCube will thus be sensitive to subtle features in the
time structure of the neutrino signal.

\subsection{SASI Sloshing and Spiral Modes and Neutrino-Emission 
Modulations}

Large-scale, large-amplitude oscillations of the stalled shock 
and postshock
layer associated with SASI motions cause modulations
of the mass-accretion flow onto the newly formed neutron star.
These accretion variations 
lead to quasi-periodic fluctuations of the neutrino emission,
which can well be detected with IceCube
for a Galactic supernova~\cite{Lundetal2010}. 
This phenomenon was first observed in 2D simulations
with RbR+ neutrino transport~\cite{MarekJankaMueller2009} as well
as multidimensional, multi-angle neutrino 
transport~\cite{Ottetal2008,Brandtetal2011}.

However, doubts about the physical reality of the phenomenon
were expressed because in 2D the imposed axial symmetry
constrains the SASI sloshing to the axial
direction and the accretion flow occurs mostly close to the
equatorial plane, potentially exaggerating the asymmetry.
Such a geometrical constraint does not exist in 3D, where
in addition symmetry-breaking ($m=1$) spiral modes can occur
due to a superposition of phase-shifted bipolar SASI 
oscillations in different directions
\cite[e.g.,][]{BlondinMezzacappa2007,Iwakamietal2009,Fernandez2010,Hanke2013}.
Angular momentum in the 
collapsing stellar core can accelerate the growth of this
spiral SASI mode~\cite{YamasakiFoglizzo2008}. 

Tamborra et al.~\cite{Tamborra2013,Tamborra2014a} analyzed the 3D 
simulations of the Garching group for neutrino-emission
asymmetries. In order to evaluate the computational models 
for the observable neutrino signals from different viewing 
directions, they integrated the neutrino emission over the 
whole visible hemisphere for all observer positions. In 
this post-processing of the numerical data, they applied
a method introduced in Ref.~\cite{MuellerE2012}
to take into account limb darkening and projection effects.
This procedure corrects the artificial enhancement of 
small-scale emission variations associated with the use of
RbR+ neutrino transport in the supernova simulations.

During phases of SASI sloshing and spiral motions the
emission of $\nu_e$ and $\bar\nu_e$, whose production
dominates in the hot accretion flows, exhibits 
synchronous time modulations with a spatial correlation.
On a somewhat lower level such emission variations can
be found also for heavy-lepton neutrinos ($\nu_x$).
The amplitude of these fluctuations can be up to 20\% 
of the direction-averaged luminosities of 
$\nu_e$ and $\bar\nu_e$ and up to about 5\% for $\nu_x$.
The mean energies of the radiated neutrinos of all species
change by about 1\,MeV, also in phase with the luminosity 
variations. The typical frequency of the fluctuations
is 50--100\,Hz and shows up as a strong peak in the 
power spectrum of the neutrino-detection rate. This
frequency is closely linked to the inverse of the 
duration of the SASI cycle, $\tau_\mathrm{cyc}$
(Eq.~\ref{eq:cycletime}), which can be expressed in terms
of the shock radius and of the neutron-star radius and mass
as~\cite{Mueller2014}
\begin{equation}
\tau_\mathrm{cyc} \approx (10\,...\,20)\,\mathrm{ms}\, 
\left(\frac{R_\mathrm{s}}{100\,\mathrm{km}}\right)^{\! 3/2}
\left(\frac{M_\mathrm{NS}}{1.5\,M_\odot}\right)^{\! -1/2}
\ln\left(\frac{R_\mathrm{s}}{R_\mathrm{NS}}\right)\,.
\label{eq:tsasi}
\end{equation}
Since the growth of the SASI is favored during periods
of shock retraction 
(cf.~Sects.~\ref{sec:sasigrowth} and~\ref{sec:sasirelevance}), 
a measurement of the duration and frequency of SASI-induced
time variations in the neutrino signal from a 
Galactic supernova would yield information about the shock 
radius and its evolution prior to the onset of the explosion.

The neutrino-emission variations predicted by the 3D models would
be well detectable with IceCube for a stellar death in the Milky
Way, in particular from observer positions close to the plane of
the SASI activity, where the fluctuation amplitude is largest.
But even for observers outside of this
plane, which is not fixed but can move and differ between
different episodes of SASI activity, there is a promising 
perspective for detection~\cite{Tamborra2013,Tamborra2014a}.  

In evolution phases without SASI shock motions and in 
models with dominant convective activity in the postshock 
layer, the neutrino luminosities still exhibit 
fluctuations caused by the convective variations of the mass
accretion by the nascent neutron star. However, these 
luminosity fluctuations look similar from all observer 
directions, and they are less regular and have smaller
amplitudes of only some percent, associated with a 
broader power distribution in the Fourier space.
IceCube will still be able to detect these signal
features, albeit only for a supernova at a distance
up to a few kpc~\cite{Lund2012}.

\begin{figure}[t!]
\includegraphics[width=14cm]{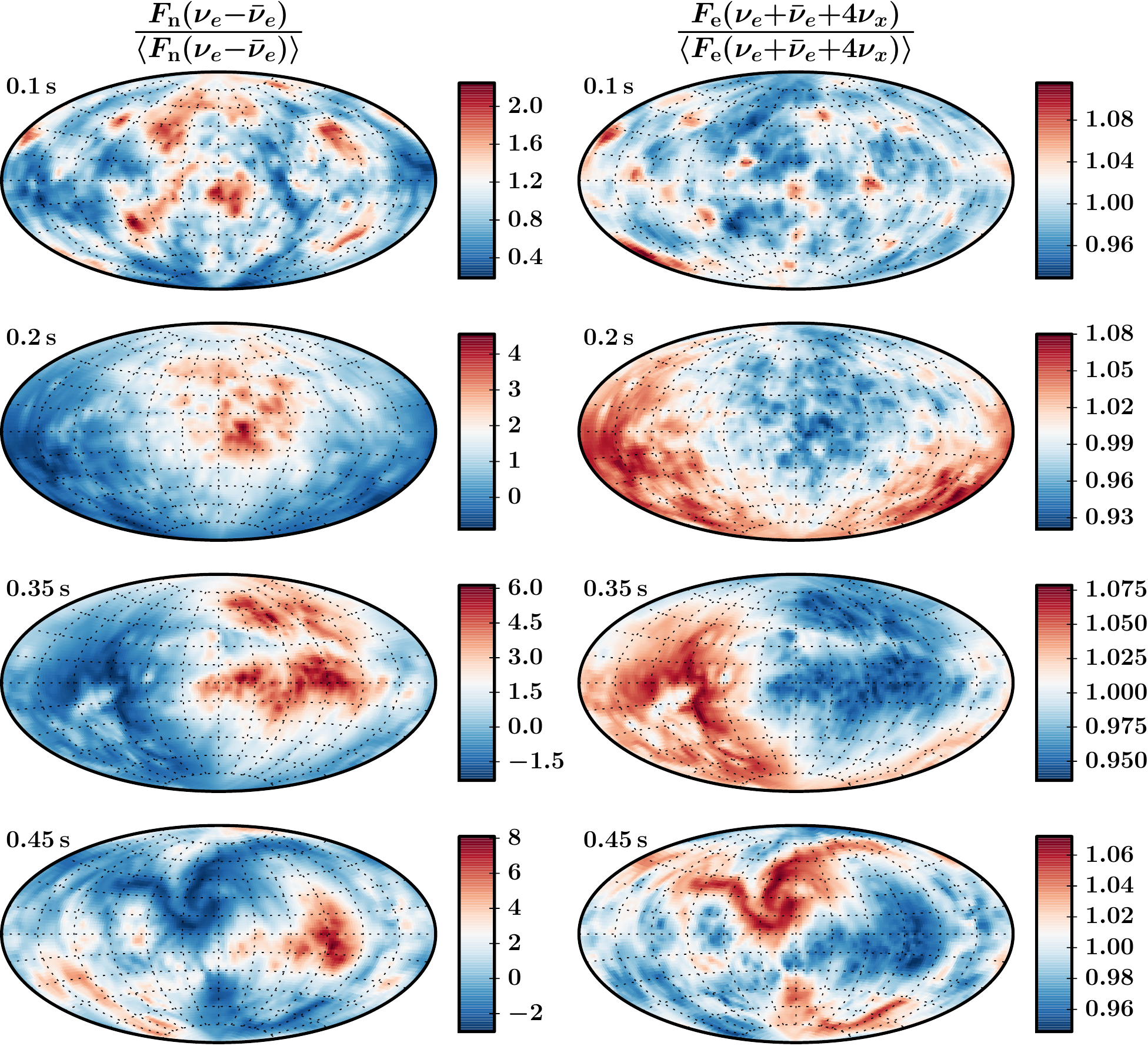}
\caption{Evolution of the LESA in a 3D simulation of a
9.6\,$M_\odot$ progenitor star. The simulation was
performed with an axis-free Yin-Yang grid. Each panel
shows an all-directions ($4\pi$) image. 
The plots in the left column present the
local lepton-number flux densities ($\nu_e$ minus $\bar\nu_e$),
normalized by their average over the whole sphere, at four
different times. The right column shows the total
neutrino-energy flux densities ($\nu_e$ plus $\bar\nu_e$ plus
the contributions from all four heavy-lepton neutrinos),
again normalized by the average value.
From the initial higher-order multipolar pattern
a clear dipolar asymmetry develops within $\sim$200\,ms.
The dipole direction remains nearly stationary. At
$t \gtrsim 350$\,ms a strong quadrupolar component
appears. While local maxima and minima of the
lepton-number flux can exceed the average value by
factors of several, the hot and cold spots of the
total energy flux reach peak values that deviate from
the direction-averaged energy flux by only a few percent.}
\label{fig:lesaevol}
\end{figure}

\begin{figure}[t!]
\includegraphics[width=15cm]{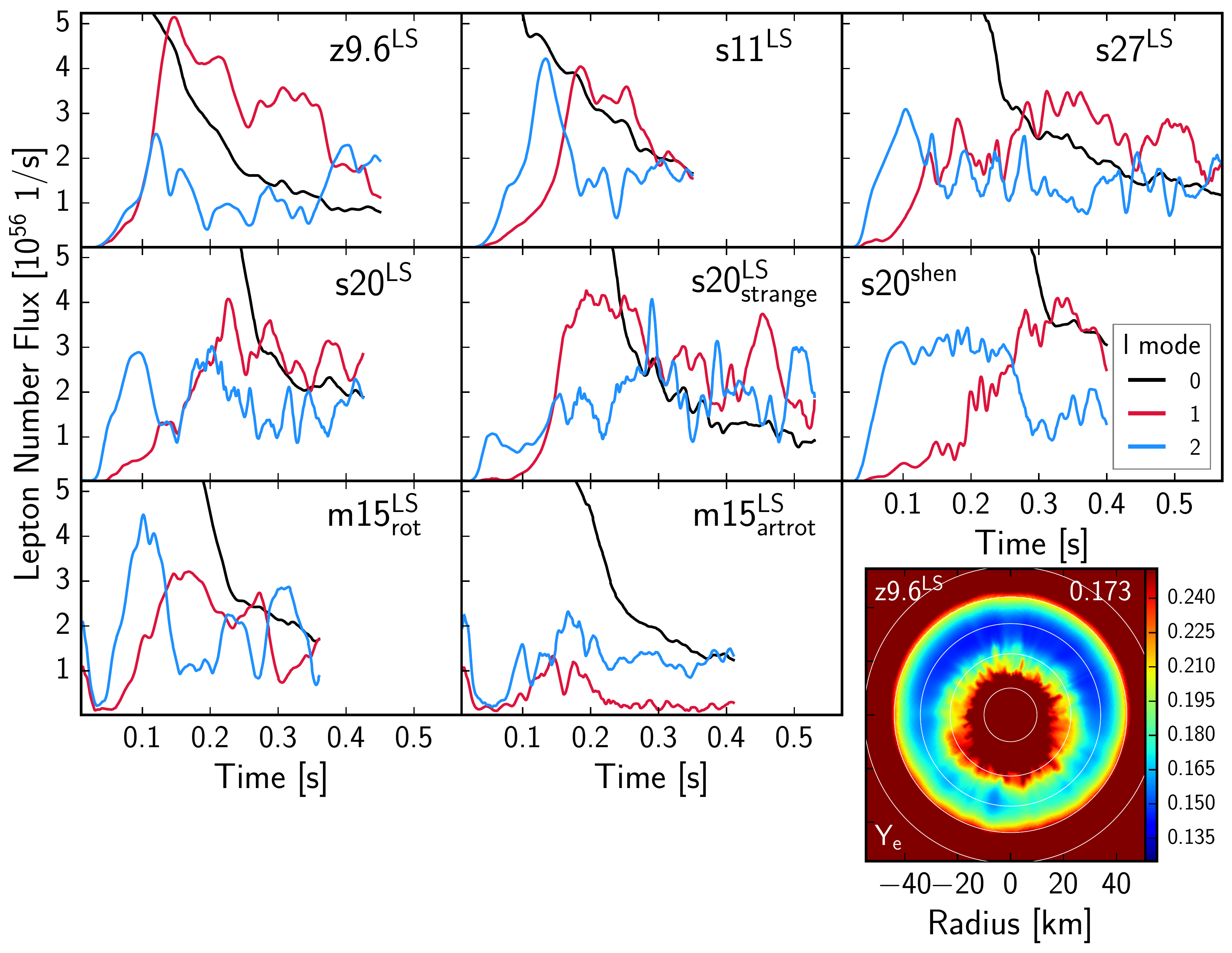}
\caption{Evolution of the lepton-number emission as
function of post-bounce time for eight 3D simulations
for nonrotating 9.6, 11.2, 27, and 20\,$M_\odot$ progenitors
(with different nuclear equations of state and 
different prescriptions of the neutrino opacities)
and for a 15\,$M_\odot$ star with slow and fast rotation
(from top left to bottom right, as labeled). Each panel 
shows the overall lepton-number flux (monopole of the 
angular distribution; black curve), and the power of the
dipolar (red curve) and quadrupolar components (blue curve).
While the monopole declines along with the contraction of
the proto-neutron star and the progenitor-dependent
decrease of the mass-accretion rate, all cases show the
development of a strong dipole with similar growth
behavior but considerable variation of the growth time
scale. The panel in the lower right corner displays the 
electron-number fraction ($Y_e$; electrons per nucleon) 
at around the time of the dipole maximum in 
the proto-neutron star of the 9.6\,$M_\odot$ model
as a representative case. The cross-sectional plane
contains the dipole axis with excess $\nu_e$ emission 
in the downward direction. The white circles correspond
to contours for densities of $10^{14}$, $10^{13}$,
$10^{12}$, $10^{11}$, and $10^{10}$\,g\,cm$^{-3}$
(from the center outward). The bluish ring is an 
asymmetric low-$Y_e$ layer interior to the neutrinosphere,
which partly overlaps with the convective shell inside
of the proto-neutron star. 
(Figure courtesy of Georg Stockinger)}
\label{fig:lesaampl}
\end{figure}

\subsection{Lepton-number Emisson Asymmetry (LESA)}

The 3D simulations of the Garching group also revealed a 
stunning, new, and unexpected neutrino-hydrodynamics
phenomenon that develops
in all models simulated with energy-dependent,
three-flavor neutrino transport handled by the
\textsc{Prometheus-Vertex} code. It was first described by
Tamborra et al.~\cite{Tamborra2014b}, who discovered a dipolar
asymmetry of the lepton-number emission in the neutrino
data, which they termed LESA: Lepton-number Emission
Self-sustained Asymmetry.

\subsubsection{LESA Phenomenology}

Within roughly 200\,ms after core
bounce a large hemispheric asymmetry of the lepton-number
emission builds up from an initial, high-order multipolar 
pattern (\textbf{Figure~\ref{fig:lesaevol}}). In fact, the 
lepton-number ($\nu_e$ minus $\bar\nu_e$) flux develops a 
dominant dipole amplitude that can become stronger than
the monopole. It implies that the newly formed neutron star
loses its lepton number predominantly in one hemisphere,
i.e., the $\nu_e$ number flux clearly exceeds the $\bar\nu_e$
number flux on one side whereas the excess is much smaller
(or even the $\bar\nu_e$ emission stronger) on the other side.

This dipole asymmetry persists for hundreds of milliseconds
with its amplitude, after passing the maximum, following 
the gradual decay of the monopole
(\textbf{Figure~\ref{fig:lesaampl}}). The dipole direction
changes only slowly in models where convective mass 
motions determine the nonradial flows in the postshock layer, 
i.e.\ it drifts on time scales much longer than the typical 
convective turnover time scale, accounting for an angular velocity
of up to $\sim$90$^\circ$ over several hundred milliseconds.
During phases with violent SASI spiral motions such a long-time
drift can be superimposed by a wobbling of the LESA direction
by up to some 10$^\circ$ with the high frequency 
(Eq.~\ref{eq:tsasi}) of the SASI
spiraling~\cite{Tamborra2014b}.

The LESA dipole direction and the SASI shock-deformation 
vector are uncorrelated and not causally connected. The
characteristic neutrino-emission properties of SASI and LESA
differ fundamentally~\cite{Tamborra2014a,Tamborra2014b}.
SASI asymmetries and time-modulations are synchronized
between all neutrino species and can reach 10--20\% of the 
total energy flux. In contrast, the amplitude of the 
lepton-number flux
dipole can even exceed the monopole, and the $\nu_e$ and 
$\bar\nu_e$ emission maxima peak in opposite hemispheres.
Different from these neutrinos, whose individual hemispheric
flux asymmetries can be as high as 20\%, heavy-lepton neutrinos 
($\nu_x$) are affected by the LESA effect only on a minor
level of order percent. The total energy flux (summed over
all neutrino species) therefore exhibits directional 
variations and a dipole amplitude on the level of several
percent with its maximum pointing opposite to the 
lepton-number emission dipole 
(\textbf{Figure~\ref{fig:lesaevol}}, right column).

\subsubsection{LESA Origin}
\label{sec:lesaorigin}

The emission asymmetry of the LESA phenomenon
originates mostly from the 
convection layer inside of the proto-neutron star. Up to
the neutrinosphere about three quarters of the final
lepton-emission dipole have built up, and another quarter
of the total effect is added in the accretion layer that 
surrounds the neutrinosphere~\cite{Tamborra2014b}.

The emission dipole is mirrored by a pronounced 
hemispheric asymmetry of the electron distribution in 
a thick shell just inside of the neutrinosphere: in the
hemisphere of the stronger $\nu_e$ emission the electron 
fraction ($Y_e$, which is the number of electrons per baryon) 
is considerably higher than on the opposite side (see
\textbf{Figure~\ref{fig:lesaampl}}, where the cross-sectional
panel shows this shell bounded by the density contours
for $10^{11}$\,g\,cm$^{-3}$ and roughly 
$10^{13}$\,g\,cm$^{-3}$).
This asymmetry is a consequence of stronger convection 
in the proto-neutron star on the side of the higher $Y_e$
and dominant $\nu_e$ emission. The correspondingly enhanced 
convective transport of lepton number carries electrons
from the lepton-rich, deep core (where $Y_e > 0.25$) to the 
more deleptonized layer ($Y_e < 0.19$) enclosed by the 
neutrinosphere. 

Despite the large hemispheric imbalance of $Y_e$ the
density and pressure distributions remain spherical (in
the absence of rotation) as dictated by the 
monopole-dominated gravitational potential. Since the
pressure of the nuclear medium is a function of density, 
temperature, and electron fraction, $P = P(\rho,T,Y_e)$,
the nonspherical distribution of $Y_e$ must be compensated
by a small asphericity of the temperature (and entropy).
Regions with high $Y_e$ must be cooler. This explains why
the total neutrino-energy flux is higher in the hemisphere
opposite to the lepton-emission maximum 
(\textbf{Figure~\ref{fig:lesaevol}}, right column).

This interior effect is amplified by an exterior
feedback cycle. Since $\bar\nu_e$ are radiated with 
somewhat harder spectra than $\nu_e$, the neutrino-energy
deposition by $\nu_e$ and $\bar\nu_e$ absorption in the
gain layer is stronger
in the hemisphere of relatively higher $\bar\nu_e$ emission
(i.e., on the side opposite to the LESA dipole direction).
The stronger heating pushes the accretion shock to a larger
radius, thus creating a global dipolar deformation of the 
shock that persists (on average)
as a long-time phenomenon, overlain by short-time
variations associated with convective and SASI activity in
the gain layer. Due to the dipolar deformation the accretion
shock deflects the incoming mass-accretion flow and channels
it preferentially to the hemisphere of the smaller shock 
radius. This effect enhances the mass accretion of the neutron
star and therefore the inflow of fresh lepton number on the 
side of the higher $Y_e$, thus further adding to the $Y_e$ 
asymmetry and amplifying the lepton-number emission asymmetry.
Tamborra et al.~\cite{Tamborra2014b} speculate that this accretion
asymmetry could function as a self-sustaining mechanism
that stabilizes the LESA dipole during the accretion phase.

\subsubsection{LESA Sensitivity to Model Variations}

\textbf{Figure~\ref{fig:lesaampl}} displays the development
of LESA in a set of 3D simulations for different progenitors,
varied microphysics, and different rotation rates.
The plots show the time evolution of the power in the 
monopole, dipole, and quadrupole components ($l = 0,\,1,\,2$)
of the spherical harmonics decomposition of the lepton-number
emission, defined by
\begin{equation}
A_l = \left(\sum^{l}_{m=-l}\left|\int_{\Omega}
\mathrm{d}\Omega \,\,
Y_{lm}^{*}(\theta,\phi) \,
r^2 \left[ F_{\mathrm{n},\nu_e}(r,\theta,\phi)
- F_{\mathrm{n},\bar\nu_e}(r,\theta,\phi) \right]
\right|^{2} \right)^{\! 1/2} ,
\label{eq:lesapower}
\end{equation}
where $F_{\mathrm{n},\nu_i}(r,\theta,\phi)$ is the local
(radial) number flux density of neutrino species $\nu_i$.
The monopole is the total lepton-number flux and
the dipole is defined in the same way as introduced in
Ref.~\cite{Tamborra2014b}:
$A_\mathrm{monopole} + A_\mathrm{dipole}\cos\vartheta = 
A_0 + A_1\cos\vartheta$ in coordinates aligned with the 
dipole direction.

The monopole decays along with the decline of the mass-accretion
rate. The quadrupole grows faster than the dipole in all
models and both appear with the onset of convection 
in a Ledoux-unstable layer inside of the neutron star.
In all cases the dipole power begins to increase
quasi-exponentially at $t \gtrsim 100$\,ms after bounce,
saturates at a similar magnitude and dominates
the quadrupole component and even the monopole at some 
point, except in the rapidly rotating model
m15$_\mathrm{artrot}^\mathrm{LS}$. Rotation suppresses
the growth of the dipole; model m15$_\mathrm{rot}^\mathrm{LS}$ 
with less angular momentum reveals this effect less strongly. 

The growth of the dipole takes place during a phase of
rapid contraction and deleptonization of the proto-neutron
star that is a consequence of its fast gain of mass 
associated with the high mass-accretion rate early after
bounce. A connection between neutron-star contraction and
LESA growth is also suggested by model
s20$_\mathrm{strange}^\mathrm{LS}$, where due to higher $\nu_x$
emission because of a reduced neutrino-nucleon scattering
opacity~\cite{Melson2015b} the neutron star contraction
and the dipole growth are faster than in the 
reference model s20$^\mathrm{LS}$. Inversely, model
s20$^\mathrm{shen}$ was computed with a stiffer nuclear
equation of state, which slows down the contraction of
the nascent neutron star. Indeed, in this simulation the
dipole growth is delayed and its climb to maximum
strength takes longer.

\subsubsection{Explanation of LESA}

The LESA phenomenon is not well understood yet and,
in particular, the development of a dominant dipole 
asymmetry still needs to be explained. 

Tamborra et al.~\cite{Tamborra2014b}
reasoned that the two opposite hemispheres ``communicate''
by the accretion asymmetry around the neutron star
(cf.~Sect.\ref{sec:lesaorigin}). A detailed analysis,
however, reveals that also in the convection layer inside
of the proto-neutron star there is an effective large-scale
flow between the two
hemispheres, which overlies the more local
mass motions associated with the small-scale pattern of
convective cells~\cite{Stockinger2015}. 

The quasi-exponential amplification of the LESA asymmetry
might be connected to the fact that convection in the 
proto-neutron star according to the Ledoux criterion,
\begin{equation}
C_\mathrm{Ledoux} = 
\left(\frac{\partial\rho}{\partial s}\right)_{\!\! P,Y_\mathrm{lep}}
\frac{\mathrm{d}s}{\mathrm{d}r} +
\left(\frac{\partial\rho}{\partial Y_\mathrm{lep}}\right)_{\!\! P,s}
\frac{\mathrm{d}Y_\mathrm{lep}}{\mathrm{d}r} > 0 \ \ 
\mathrm{for\ instability}
\label{eq:ledoux}
\end{equation}
($Y_\mathrm{lep}$ is the electronic lepton fraction and
$s$ the entropy per nucleon),
depends on entropy and lepton-number gradients. In this 
context it is important that 
$\left(\partial\rho/\partial Y_\mathrm{lep}\right)_{P,s}$
changes its sign from negative to positive for values of 
$Y_\mathrm{lep}$ below a critical limit. This critical
threshold is a function of density and temperature and 
increases for higher $\rho$ and $T$. As the proto-neutron
star deleptonizes and contracts, the threshold is eventually
passed, in which case a negative gradient of 
$\mathrm{d}Y_\mathrm{lep}/\mathrm{d}r$ becomes stabilizing
instead of destabilizing. This effect damps convective 
activity. As a consequence, electrons are less efficiently 
dragged upward from the lepton-rich central core, while at
the same time outward neutrino diffusion (which becomes
faster than convective transport near the outer edge of
the convection zone) still carries
away lepton number and maintains the ongoing deleptonization
of the convective shell.
With the decreasing lepton fraction the stabilizing influence
of the second term in the Ledoux criterion will further
increase. This defines an amplifying feedback cycle that
could be the underlying reason for the exponential growth
of the lepton-emission asymmetry, instigated by a 
sufficiently large seed perturbation.

LESA would thus manifest itself as a neutrino-hydrodynamics
instability, in which convective and neutrino transport
in combination determine the large-scale flow dynamics 
within global, nonspherical modes that encompass a major 
fraction of the volume of supernova cores.

\subsubsection{Is LESA a Numerical Artifact?}

All facts reported in the preceding sections seem to
provide support for LESA being a physics phenomenon and
not a numerical artifact, e.g.\ as a consequence of the RbR+ 
transport treatment applied in the 3D simulations
performed at Garching. However, a causal connection between 
LESA and RbR(+) transport appears unlikely, because the
RbR(+) approximation tends to produce localized, small-scale
extrema, and it is difficult to imagine how this could help
assembling a global dipole asymmetry.
LESA is certainly not linked to the use of a polar
coordinate grid (with its well-known axis artifacts),
because the dipole direction differs from model to 
model (with its beginning not being correlated with
the direction of the polar axis), and because a 
dipolar lepton-emission asymmetry also develops in 
simulations with an axis-free Yin-Yang grid (e.g.\ in
models z9.6$^\mathrm{LS}$ and m15$_\mathrm{rot}^\mathrm{LS}$ 
of \textbf{Figures~\ref{fig:lesaevol}}
and~\textbf{\ref{fig:lesaampl}}).

Yet, plausibility and consistency 
is not evidence. Independent confirmation
by other groups is needed and a theoretical framework must
be developed that assembles the different pieces of the
puzzle described above into a consistent picture.
Indeed, the presence of a hemispheric $Y_e$ asymmetry
in the proto-neutron star and of a lepton-number emission
dipole with a size of order the monopole magnitude was 
also observed in 3D simulations with a neutrino-leakage
scheme (Evan O'Connor, private communication 2014). 
Also the 15\,$M_\odot$ 3D explosion model of the 
Oak Ridge group~\cite{Lentz2015}, which applied
a flux-limited diffusion solver for the neutrino transport, 
exhibits a lepton-number emission dipole. This feature
grows at the same
time as in the Garching models and reaches peak amplitudes
of more than $10^{56}$\,s$^{-1}$ (Eric Lentz, private
communication 2015), which is again in the ballpark of
the dipole amplitude in the Garching simulations (see 
\textbf{Figure~\ref{fig:lesaampl}}). However, in the 
Oak Ridge model the monopole remains much larger 
and does not decay below $4\times 10^{56}$\,s$^{-1}$ until
450\,ms after bounce, which is a clear difference to the
Garching simulations but might be connected to the
use of different transport treatments and neutrino
opacities by the two groups.

Despite being assuring, the caveat is that all of these
3D simulations were performed with RbR(+) approximations
for the neutrino transport. Stronger evidence for the
physical nature of LESA requires its confirmation by
models with true multidimensional transport treatments.

\subsubsection{Consequences of LESA}

The LESA phenomenon implies a larger excess of the $\nu_e$ 
emission compared to the $\bar\nu_e$ emission in one
hemisphere than in the other. This asymmetry persists
until the onset of the explosion and beyond (see 
models z9.6$^\mathrm{LS}$ and s20$_\mathrm{strange}^\mathrm{LS}$,
\textbf{Figures~\ref{fig:lesaevol}}
and~\textbf{\ref{fig:lesaampl}}). It leads to different
neutron-to-proton ratios (and $Y_e$) in the supernova ejecta 
expelled in different directions, because $Y_e$ in the expanding
neutrino-heated matter is set by the competition of $\nu_e$
and $\bar\nu_e$ absorptions on neutrons and protons, respectively.
The consequences for supernova nucleosynthesis remain to 
be explored.

It is presently not known how long into the proto-neutron star
cooling evolution the strong dipole component of the neutrino 
emission will survive. If a stable dipole radiates the total
gravitational binding energy of the newly formed neutron star
with an asymmetry of the order of several percent, the 
corresponding recoil acceleration could explain the observed
natal kick velocities of pulsars up to more than 
1000\,km\,s$^{-1}$.
In the 9.6\,$M_\odot$ explosion (the 3D model z9.6$^\mathrm{LS}$)
we estimate a (mainly neutrino-induced) kick of 35\,km\,s$^{-1}$.
within the simulated time of 450\,ms after bounce. Though 
this is not particularly big, even such a modest kick 
magnitude would have important astrophysical implications if
it defines a lower bound because LESA is a generic phenomenon
that plays a role in all new-born neutron stars.

The LESA asymmetry is also of great importance for detailed
predictions of the observable neutrino signal from a future
Galactic supernova. While a viewing-angle dependence of
the radiated neutrino luminosities adds complexity to the 
interpretation of a measured neutrino burst, the picture
will become even more complicated if the direction-dependent
$\nu_e$-$\bar\nu_e$ flux asymmetry leads to 
self-induced neutrino-flavor conversions that vary with the
observer position~\cite{Chakraborty2015}. 	

With all these questions being barely scratched, the LESA 
defines yet another interesting territory that deserves
further exploration as simulations penetrate even deeper
into the landscape of 3D phenomena happening inside of
supernova cores.

\begin{summary}[SUMMARY POINTS]
\begin{enumerate}
\item 
First self-consistent 3D simulations 
with state-of-the-art neutrino transport and input physics
have yielded explosions and give hope that the neutrino-driven 
mechanism can ultimately be consolidated.
\item 
Multidimensional flows lead to 
enhanced neutrino-energy deposition and higher heating
efficiency, and nonradial fluid instabilities like buoyant
convection, turbulence, SASI sloshing and spiral motions,
and rotation facilitate shock revival.
\item 
A universal relation for the critical neutrino luminosity
is proposed that generalizes the critical luminosity condition
of Burrows \& Goshy~\cite{BurrowsGoshy1993} and accounts for 
the influence
of multidimensional effects like turbulence and rotation on the 
neutrino heating required for explosion.
\item 
Time-variations of the mass-accretion rate of the new-born
neutron star associated with large-amplitude SASI sloshing and 
spiral motions lead to quasi-periodic modulations of the 
neutrino emission, whose measurement in the case of a
future Galactic supernova would provide important information
about the shock dynamics prior to explosion.
\item 
The first 3D supernova models led to the discovery of
a lepton-number emission asymmetry (LESA) with a large
dipolar component, which ---if real and not a numerical
artifact--- may constitute a new kind of neutrino-hydrodynamical
instability and would have important consequences for 
neutron-star kicks as well as supernova nucleosynthesis and 
neutrino detection.
\end{enumerate}
\end{summary}

\begin{issues}[FUTURE ISSUES]
\begin{enumerate}
\item  {\em Better resolution.}
Current self-consistent, high-fidelity supernova
simulations are performed with severely constrained numerical
resolution. Although convergence tests suggest that this is 
sufficient to track the basic properties of nonradial flows
including turbulence effects 
\cite{Radice2015,Radice2016,Abdikamalov2015},
subtle differences may depend on
higher resolution, in particular when neutrino physics is
self-consistently included. This demands 3D simulations with
refined numerical zoning.
\item  {\em Code comparisons.}
Results of 2D and 3D supernova simulations published
by different groups exhibit significant quantitative and partly
even qualitative differences. The origin of these differences
can be linked to numerical aspects of the applied codes or
differences in the input physics. Although a careful assessment
of these possibilities does not reveal obvious contradictions,
tracing the differences back to their actual roots requires
detailed and careful direct comparisons.
\item  {\em Improved and consistent microphysics.}
Current 3D simulations are less prone to blow up than 2D
models and explode or fail marginally. Relatively small
differences in the microphysics, in particular in the
neutrino opacities, can be decisive for
success~\cite{Melson2015b}. A careful assessment of the
current treatment of neutrino interactions is needed,
in particular also in the subnuclear regime, which is
most relevant for the development of the explosion.
\item  {\em Progenitor Asymmetries.}
The potential relevance of nonspherical perturbations in
the convective burning shells of pre-collapse stars has been 
pointed out for a long time 
\cite[e.g.][and references therein]{ArnettMeakin2011},
and recent toy-model/parametric studies 
have demonstrated that such perturbations can make the
difference between success or failure in marginal
cases~\cite{CouchOtt2013,Mueller2015,CouchChat2015}.
Self-consistent simulations of the pre-collapse, infall,
and post-bounce phases are needed to clarify the importance
of burning-shell asymmetries for more realistic conditions
and in dependence of the progenitor star.
\item  {\em Multidimensional neutrino transport.}
The RbR+ approximation currently employed in
self-consistent 3D supernova simulations needs to be tested
against truly multidimensional neutrino transport treatments like
M1 closure schemes~\cite{Kuroda2015} or multi-angle 
Boltzmann-solvers~\cite{Nagakura2014,Sumiyoshi2015}. Time-dependent
simulations with such improvements and sufficiently good
resolution will require exascale computing.
\item  {\em Consequences of rotation.}
Rotation deserves more investigation in self-consistent
supernova models including neutrino transport. Current 
parametric studies suggest that the growth of spiral
SASI modes is enhanced even for moderate rotation. These
symmetry-breaking modes can foster 
explosions~\cite{Nakamura2014,Iwakami2014} and
redistribute angular momentum in the supernova core,
which has a bearing on the origin of neutron-star 
spins \cite{BlondinMezzacappa2007,Fernandez2010,Guilet2014,Kazeroni2016}.
\item  {\em Explosion energies.}
The question how 3D flow instabilities facilitate the onset of
the supernova explosion is complementary to the question how
3D hydrodynamics affects the development of the explosion
energy. With respect to the latter problem interesting 
effects favoring higher energies in 3D than in 2D are
suggested by recent simulations~\cite{Melson2015a,Mueller20152d3d}.
Reliable predictions of the supernova energetics require
long-time simulations with detailed neutrino transport
that follow the evolution for seconds beyond the onset
of the explosion. Well-resolved calculations of this phase
of simultaneous mass acceretion and outflow will pose a 
major computational challenge.
\end{enumerate}
\end{issues}

\section*{DISCLOSURE STATEMENT}
The authors are not aware of any affiliations, 
memberships, funding, or financial holdings that
might be perceived as affecting the objectivity of this 
review.

\section*{ACKNOWLEDGMENTS}
We thank Ewald M\"uller and Georg Raffelt 
for illuminating discussions,
Evan O'Connor and Eric Lentz for communicating their
LESA results to us, Aaron D\"oring (MPA), Elena Erastova, and
Markus Rampp (MPCDF) for
visualization support of the 3D simulations, and Georg
Stockinger for \textbf{Figure~\ref{fig:lesaampl}}. This research
was supported by the Deutsche Forschungsgemeinschaft through
grant EXC 153 and by the European Research Council through
grant ERC-AdG No.\ 341157-COCO2CASA. 
We acknowledge computing time from
the European PRACE initiative on SuperMUC (GCS@LRZ, Germany),
Curie (GENCI@CEA, France), and MareNostrum (BSC-CNS, Spain).
Postprocessing of simulation results was done on Hydra of 
the Max Planck Computing and Data Facility (MPCDF).

 %

%
\bibliography{JankaReferences-file}
\bibliographystyle{ar-style5}
\end{document}